\documentclass[fleqn,usenatbib]{mnras}
\usepackage{amsmath,amssymb,graphicx}
\usepackage{newtxtext,newtxmath}
\usepackage[T1]{fontenc}

\DeclareRobustCommand{\VAN}[3]{#2}
\let\VANthebibliography\thebibliography
\def\thebibliography{\DeclareRobustCommand{\VAN}[3]{##3}\VANthebibliography}

\usepackage{microtype}
\usepackage{enumerate}

\title[Barred galaxies in EAGLE]{The evolution of barred galaxies in the EAGLE simulations}

\author[M. K. Cavanagh et al.]{
Mitchell K. Cavanagh,$^{1}$\thanks{E-mail: mitchell.cavanagh@icrar.org (MKC)}
Kenji Bekki,$^{1}$
Brent A. Groves$^{1,2}$
and Joel Pfeffer$^{1}$
\\
$^{1}$International Centre for Radio Astronomy Research, The University of Western Australia, 7 Fairway, Crawley, WA 6009, Australia\\
$^{2}$Research School of Astronomy and Astrophysics, Australian National University, Mt Stromlo Observatory, Weston Creek, ACT 2611, Australia\\
}

\date{Accepted XXX. Received YYY; in original form ZZZ}

\pubyear{2015}

\begin{document}
\label{firstpage}
\pagerange{\pageref{firstpage}--\pageref{lastpage}}
\maketitle

\begin{abstract}
We study the morphologies of 3,964 galaxies and their progenitors with $M_\star > 10^{10} M_\odot$ in the reference \textsc{EAGLE} hydrodynamical simulation from redshifts $z=1$ to $z=0$, concentrating on the redshift evolution of the bar fraction. We apply two convolutional neural networks (CNNs) to classify 35,082 synthetic g-band images across 10 snapshots in redshift. We identify galaxies as either barred or unbarred, while also classifying each sample into one of four morphological types: elliptical (E), lenticular (S0), spiral (Sp), and irregular/miscellaneous (IrrM). We find that the bar fraction is roughly constant between $z = 0.0$ to $z = 0.5$ (32\% to 33\%), before exhibiting a general decline to 26\% out to $z = 1$. The bar fraction is highest in spiral galaxies, from 49\% at $z = 0$ to 39\% at $z = 1$. The bar fraction in S0s is lower, ranging from 22\% to 18\%, with similar values for the miscellaneous category. Under 5\% of ellipticals were classified as barred. We find that the bar fraction is highest in low mass galaxies ($M_\star \leq 10^{10.5} M_\odot$). Through tracking the evolution of galaxies across each snapshot, we find that some barred galaxies undergo episodes of bar creation, destruction and regeneration, with a mean bar lifetime of 2.24 Gyr. We further find that incidences of bar destruction are more commonly linked to major merging, while minor merging and accretion is linked to both bar creation and destruction.
\end{abstract}

\begin{keywords}
galaxies: bar -- galaxies: evolution -- galaxies: general
\end{keywords}


\section{Introduction}

Galactic bars are central, rectangular-shaped, internal morphological structures present in at least a 25\% to 30\% of all galaxies \citep{nair2010a,masters2011,melvin2014} and around half to more than two thirds of all spiral galaxies \citep{eskridge2000,aguerri2009,saha2018,lee2019}. The prevalence of bars throughout the Universe -- the bar fraction -- is sensitive to many factors such as environment, colour, mass and gas content \citep{lee2012,masters2012,skibba2012,vera2016,erwin2018} and is known to evolve with redshift, with a lower overall bar fraction at higher redshifts compared to the present day \citep{sheth2008,melvin2014}. Bars are important drivers of galaxy evolution \citep{cheung2013,conselice2014}. They affect many key galaxy processes including the distribution of angular momentum \citep{weinberg1985,athanassoula2005}, secular evolution \citep{kormendy2004,sheth2005,sellwood2014}, interactions with the stellar disc and surrounding dark matter halo \citep{athanassoula2002,valenzuela2003}, gas dynamics and metallicities \citep{bournaud2002,berentzen2007,fanali2015,fraser-mckelvie2019}, star formation, especially in the centres of galaxies \citep{ellison2011,lin2020}, AGN activity \citep{alonso2014,galloway2015} and quenching \citep{masters2012,kruk2018,fraser-mckelvie2020,geron2021}.

Given the evolutionary importance of bars, there is a motivation to examine bars in cosmological simulations. Studies have examined bars in several simulations, such as EAGLE \citep{schaye2015}, Illustris/TNG \citep{vogelsberger2014,pillepich2018} and NewHorizon \citep{dubois2021}. These have determined a large variation in bar fractions, ranging from nearly 40\% for EAGLE \citep{algorry2017}, 40\% to 55\% for TNG100 \citep{rosas-guevara2019,zhao2020}, to as low as 9\% in Illustris-1 \citep{peschken2019,zhou2020} and around 7 to 1\% for NewHorizon \citep{reddish2021}. Such a range in bar fractions is reflective of the inherent challenge of modelling the physical processes governing the formation and evolution of galaxies \citep{somerville2015,vogelsberger2020}.
Some studies focusing on simulated barred galaxies (e.g \citealt{emsellem2014,goz2015,okamoto2015,spinoso2017}) have utilised zoom-in simulations in order to model the physical processes important to bars, especially baryonic feedback processes \citep{zana2019} and gas dynamics, in high detail. However, such zoom-in simulations are strictly limited to analysing only a small number of galaxies, and thus lack the ability to study the impact of bars on galaxy evolution at cosmological scales. Other studies have instead examined the large-scale statistics of a population of barred galaxies in cosmological simulations, such as \textsc{EAGLE} \citep{algorry2017}.

In this work, we massively expand upon the scale of the latter approach through the use of convolutional neural networks (CNNs) to rapidly classify tens of thousands of galaxies. We study the morphological evolution of simulated galaxies in the reference EAGLE simulation from $z=1$ (lookback time of 7.93 Gyr) to $z=0$, with a focus on the bar fraction and the properties of barred galaxies. Cosmological simulations offer important insights into the impacts that bars have on galaxy evolution over cosmic time \citep{okamoto2015}, when and how bars form, whether bars form preferentially in certain galaxies over others, as well as the typical lifetimes of bars. To facilitate this analysis, we utilise CNNs to identify barred galaxies in each of the ten snapshots between $z=1$ and $z=0$.

Identifying barred galaxies, especially via visual inspection, is a difficult task, yet is well suited to automated techniques such as machine learning. Traditional methods for automatically detecting and/or classifying barred galaxies include taking a Fourier analysis of azimuthal profiles \citep{ohta1990,odewahn2004} and/or deprojected images \citep{garcia-gomez2017}, photometric decompositions \citep{reese2007,durbala2009}, ellipse fitting of isophotes \citep{abraham1999,laine2002,erwin2005,consolandi2016}, and visual inspection either by expert classifiers \citep{nair2010,buta2015} or volunteer citizen-science \citep{willett2013}.
More recently, there have been many applications of machine learning, particularly CNNs, to classify various galaxy morphologies \citep{dieleman2015,dominguezsanchez2018,zhu2019,barchi2020,martin2020,walmsley2020,vega-ferrero2021}, including galaxy bars \citep{abraham2018,cavanagh2020}. Such is the versatility of CNNs that they have been applied to a broad range of wider topics within astronomy, ranging from detecting galaxy mergers \citep{ackermann2018}, gravitational lensing \citep{schaefer2018}, exoplanets \citep{osborn2020}, classifying galaxy formation processes \citep{diaz2019}, identifying radio sources \citep{lukic2019}, to cosmological models \citep{lucie-smith2018,mustafa2019,li2021}.

For this work, we use two CNNs, each trained on g-band SDSS DR7 \citep{york2000,abazajian2009} images from the \citet{nair2010} (hereafter NA10) catalogue, to classify synthetic g-band galaxy images as either barred or unbarred, and as one of four morphological types; elliptical (E), lenticular (S0), spiral (Sp) and irregular/miscellaneous (IrrM). We refer to these CNNs as the ``bar CNN'' and ``4-way CNN'' respectively. The bar CNN is new to this work, while the 4-way CNN is based on our previous work (\citealt{cavanagh2021}, hereafter C21). We classified synthetic images of EAGLE galaxies with stellar masses above $M_\star \geq 10^{10} M_\odot$ for all ten snapshots between $z=1$ and $z=0$. The synthetic images are set to mimic the SDSS g-band imaging. It is worth noting that the bar fraction is dependent on morphological band, with bars more prevalent in the infrared \citep{eskridge2000,menendez-delmestre2007,buta2010}. However, the primary focus of this work is on g-band classifications, as it is easily comparable to surveys such as the SDSS and Galaxy Zoo \citep{lintott2011,willett2013}, and since the CNN models are trained on g-band data.

The structure of our paper is as follows. In \S 2, we briefly summarise the \textsc{EAGLE} simulations. We describe how the synthetic \textsc{EAGLE} images were generated, and how these were processed for use with the CNNs. We further outline the CNN models used to perform the classifications, briefly describe the data processing workflow, and lastly highlight how the bar CNN model was trained and evaluated. In \S 3, we present our key results on the redshift evolution of bar fractions in terms of overall samples, as well as across different mass ranges and morphological types, and briefly relate these results to observed bar fractions. We also present the evolution of morphological fractions for each of the four morphological types. In \S 4, we focus on the evolution of bars and their physical properties, especially regarding the differences in barred lenticulars and barred spirals. We examine the creation and destruction of bars and the extent to which such events are associated with galaxy merging. We also examine the caveats and limitations of our machine learning approach, and discuss the overall performance of our CNNs. Lastly, we summarise our key results in \S 5.

\section{Methods}

\begin{figure*}
\centering
\includegraphics[scale=0.26]{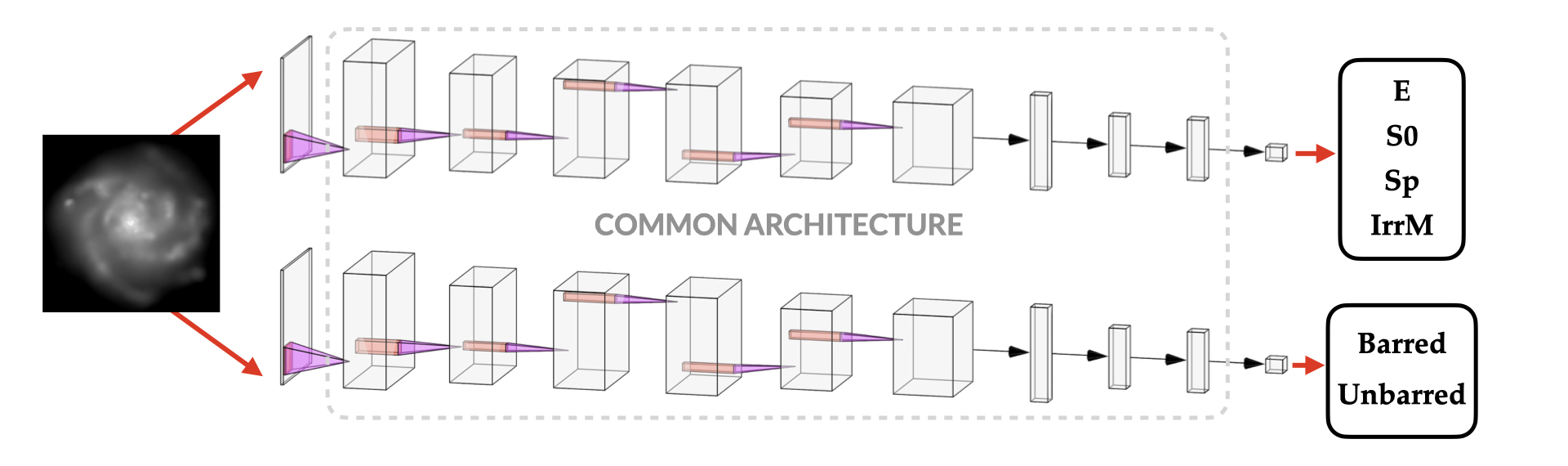}
\caption{Schematic representation of our image classification workflow. Each image is classified by two CNNs. One has been trained to classify morphological types (either E, S0, Sp or IrrM), the other to determine whether the sample is barred. Each network shares the common C2 architecture from \citet{cavanagh2021} (C21), differing only in the size of the output layer.}
\label{fig:cnn}
\end{figure*}

\subsection{The EAGLE Simulation}

The `Evolution and Assembly of GaLaxies and their Environments' (EAGLE) project \citep{crain2015, schaye2015} is a suite of cosmological hydrodynamical simulations of galaxy and evolution formation in the $\Lambda$ cold dark matter cosmogony.
The simulations adopt the cosmological parameters $\Omega_\mathrm{m} = 0.307$, $\Omega_{\Lambda} = 0.693$, $\Omega_\mathrm{b} = 0.04825$ and $\sigma_8 = 0.8288$, consistent with a \citep{planckcollaboration2014} cosmology.
The simulations were performed with a modified version of the $N$-body Tree-PM smooth particle hydrodynamics code \textsc{Gadget 3} \citep{springel2005} and include subgrid routines for radiative cooling, star formation, stellar evolution, stellar feedback, black holes (BHs) and AGN feedback (for details, see \citealt{schaye2015}).
The stellar and BH feedback parameters were calibrated such that at $z \approx 0$ the simulations match the observed galaxy stellar mass function, galaxy sizes and BH masses \citep{crain2015}.
The EAGLE simulations have been shown to broadly reproduce many properties of the evolving galaxy population, such as stellar masses and specific star formation rates \citep{furlong2015}, sizes \citep{furlong2017}, colours \citep{trayford2015, trayford2017} and cold gas properties \citep{lagos2015, crain2017}.

Bound structures (subhaloes/galaxies) are identified in the simulations using the friends-of-friends \citep[FoF,][]{davis1985} and \textsc{subfind} algorithms \citep{springel2001, dolag2009}.
Galaxies are linked between consecutive snapshots through merger trees using the D-Trees algorithm \citep{jiang2014, qu2017}.
The galaxy merger trees are available from the EAGLE public data base \citep{mcalpine2016}.

In this work we use the EAGLE `reference' simulation (Ref-L100N1504) of a periodic volume with a side length of $100$ comoving Mpc. The volume contains $1504^3$ dark matter and gas particles initially with particle masses of $m_\mathrm{dm} = 9.7 \times 10^6 M_\odot$ and $m_g = 1.8 \times 10^6 M_\odot$, respectively. The softening length is 2.66 co-moving kpc up to a maximum of 0.7 proper kpc, which is reached at $z=2.8$. For our purposes, the softening length remains fixed at 0.7 kpc since we are restricted to snapshots with $z \leq 1$. Given the constraints on the simulation parameters, we note that artificially thicker discs, as well as disc heating due to particle scattering, may hinder the formation of bars through gravitational instabilities \citep{bauer2019}, which may result in lower bar fractions compared to observations.

\subsection{Synthetic Images and Data Processing}

We first select all galaxies with stellar masses $M_\star > 10^{10} M_\odot$ at snapshots between $z=0$ and $z=1$. Such galaxies are well resolved with $\gtrsim 10^4$ stellar particles. We first orient the galaxies face on by calculating the spin vector for all stars between $2.5$ and $30$~kpc from the centre of potential of each galaxy. The radius limits ensure that the spin is dominated by the rotating disc component of each galaxy (if present) rather than central bulge regions \citep{trayford2017}.

We calculate rest-frame SDSS $g$-band luminosities for each stellar particle using the \textsc{fsps} stellar population model \citep{conroy2009, conroy2010} with a \citep{chabrier2003} initial stellar mass function, Miles spectral library \citep{sanchez-blazquez2006}, Padova isochrones \citep{girardi2000} and assuming simple stellar populations.
We assume stars with ages $<10$~Myr are fully embedded within an optically thick cloud as a simple model for dust absorption \citep{charlot2000}.

The synthetic images were generated using \textsc{SPHviewer} \citep{benitez-llambay2015}, where particles are distributed according to their local smoothing length.
We adopt a method somewhat similar to \citet{trayford2017}, by resampling the young stellar populations, from both the existing young star particles ($<100$~Myr) and star-forming gas particles, at higher resolution ($10^3 M_\odot$) based on the star-formation rate of the particles (assuming a constant formation rate). 
The resampled particles adopt the same position, metallicity and smoothing lengths as the particle from which they are sampled, while new ages are randomly assigned between 0 and 100~Myr.
Young star particles ($<100$~Myr) then adopt the local SPH smoothing length, while for older stars we recalculate a local stellar smoothing length using the nearest 64 neighbours.
This method ensures that star formation in (e.g.) the spiral arms of galaxies is adequately sampled.

The convolutional neural networks utilised in this work were designed to classify monochromatic images with $100 \times 100$ pixels of size $0.5$ kpc, corresponding to a $50$ kpc$^{2}$ image. As the CNN was trained on SDSS $g$-band images, we similarly generate $50 \times 50 \, \mathrm{kpc}^2$ rest frame $g$-band images with pixel sizes of $0.5$ kpc for the EAGLE galaxies in each snapshot. The pixel data was linearly normalised such that every pixel has a minimum and maximum values of 0 and 1 respectively. This is consistent with the normalisation applied to the data that was initially used to train the models, as per C21. It is important to ensure that inputs to a CNN are normalised, as this helps ensure optimal model convergence and improves the model's ability to generalise \citep{lecun2015,gu2018}. As such, future applications of the models require normalised inputs. We generated a total of 35,082 g-band images for 3,964 unique galaxies. Of these, 2,146 galaxies were present for all ten snapshots between $z=0$ and $z=1$. There are an average of 3,508 galaxies per snapshot.

\subsection{CNN Models}

CNNs are powerful tools that are able to identify and extract relevant features of data in a model-independent manner, relying solely on a sufficiently large training set of data with which to learn from \citep{lecun2015,goodfellow2016,gu2018,dhillon2020}. CNNs are readily applicable to domains involving image classification, and have been increasingly utilised in astronomy (see \citealt{baron2019} for a review).

This work utilises two separate CNN models, each sharing a common architecture. One is trained to classify barred or unbarred galaxies (hereafter referred to as the bar CNN), the other is trained to classify samples into one of four morphological types; E, S0, Sp and IrrM (known hereafter as the 4-way CNN). The 4-way CNN is directly based on our previous work (\citealt{cavanagh2021}, C21) – this work also describes the common C2 architecture and how the four morphological types are defined in terms of NA10 catalogue. The bar CNN is a newly developed model trained specifically for this work, and is described in \S 2.4. Both models are trained on g-band SDSS images of galaxies from the NA10 catalogue. The 4-way model from C21 has an overall accuracy of 81\%,  however the per-class accuracy for irregular/miscellaneous samples was significantly lower than for the other three classes (27\%) due to the extremely low number of irregular training samples.

Figure \ref{fig:cnn} represents the workflow for image classification. Each galaxy image is classified by the two CNNs to determine its morphological type and whether it is barred. We stress that each classification - bar and 4-way - is carried out independently, and the results of one model do not influence the other. Each model outputs an array of probabilities for each category. The final, predicted class is simply the category corresponding to the highest probability. In this context, the classification probabilities are also known as confidences, and we will use these terms interchangeably throughout the remainder of this work.

\begin{figure}
\centering
\includegraphics[scale=0.65]{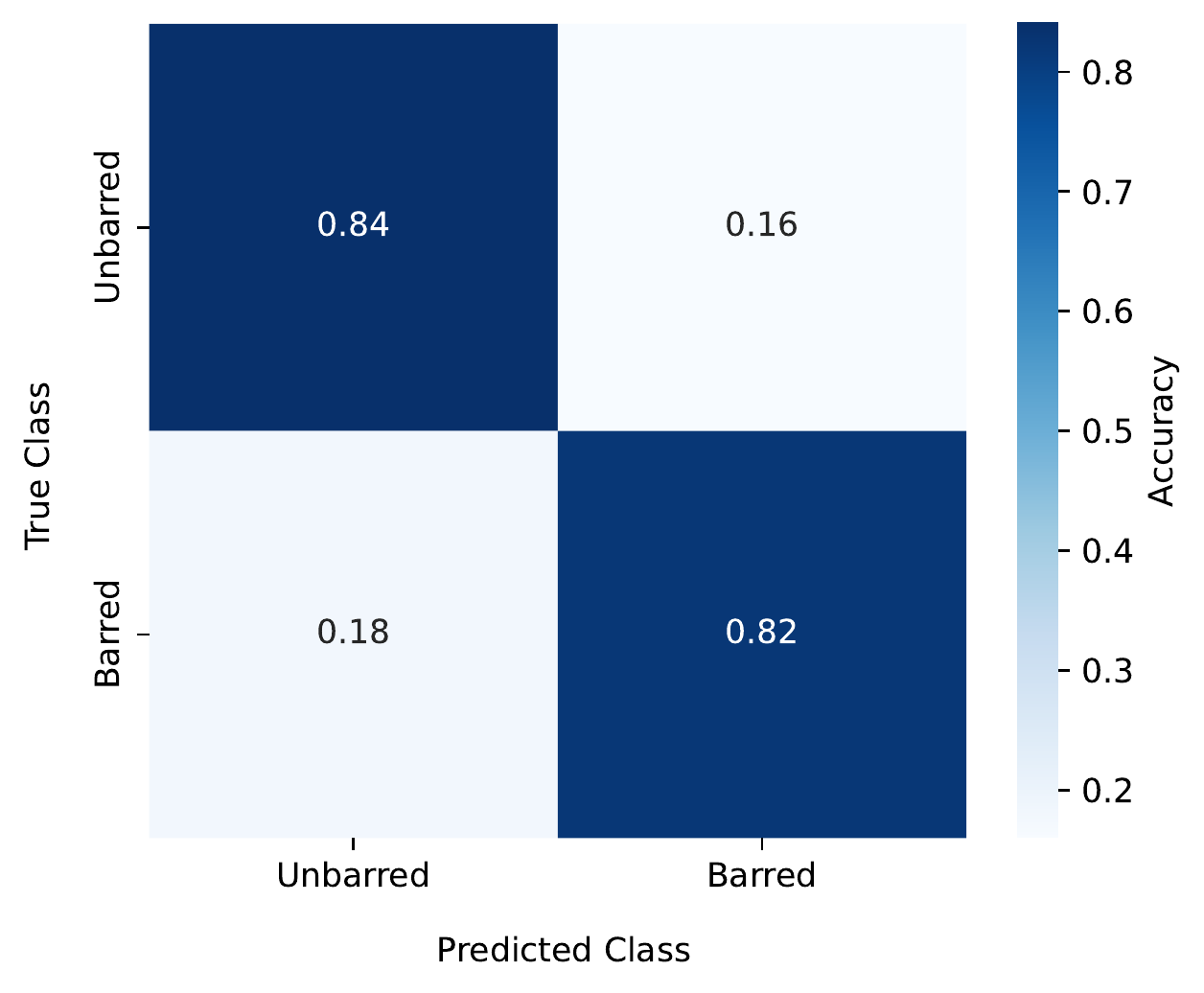}
\caption{Confusion matrix for our bar CNN, shaded according to classification accuracy. The vertical axis corresponds to the true class while the horizontal axis corresponds to the predicted class.}
\label{fig:confmat}
\end{figure}
\begin{figure*}
\centering
\includegraphics[scale=0.45]{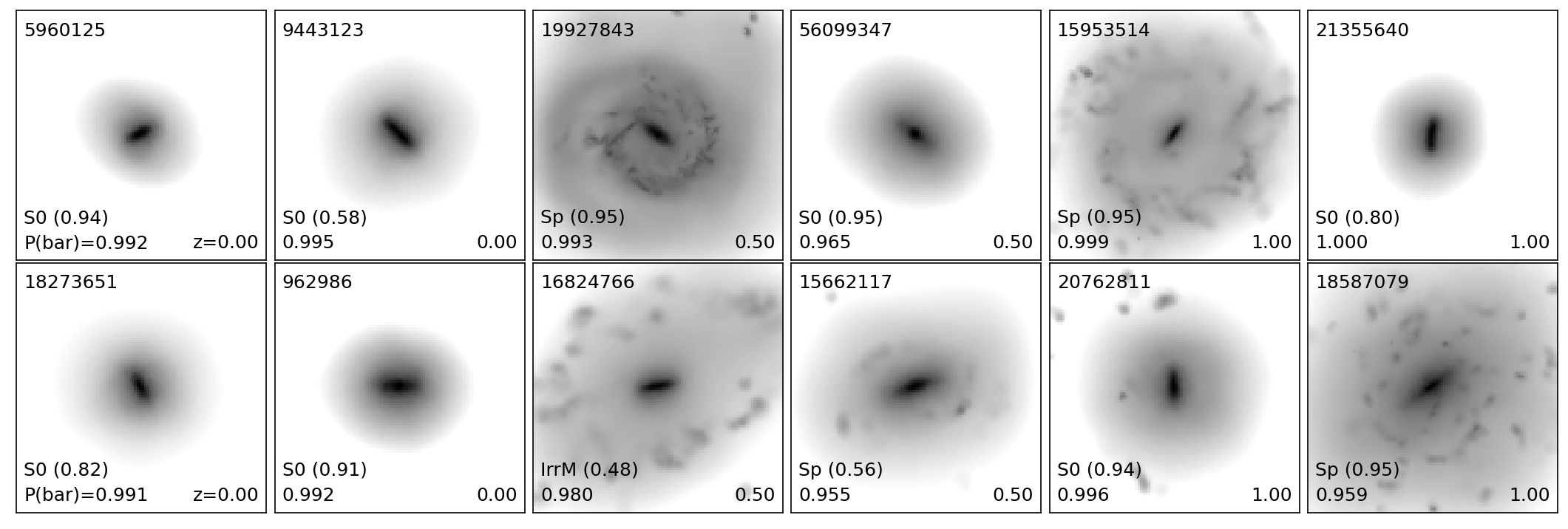}
\caption{A random selection of barred galaxies for low, medium and high redshifts, with classification confidences $P(\text{bar}) > 0.95$. Each image has the same pixel size and normalisation as what was loaded directly into the CNN, as opposed to the raw synthetic image. Each image is annotated with their \textsc{EAGLE} CATID in the top left, the CNN morphological classification (with confidence in brackets) plus the probability of being barred in the bottom left and redshift in the bottom right.}
\label{fig:exbars}
\end{figure*}

\subsection{The Bar CNN}

The bar CNN is trained to classify galaxies as either barred or unbarred. The data used to train the network consisted of g-band SDSS DR7 images of galaxies sourced from the NA10 catalogue of morphological classifications. Although the NA10 catalogue delineates bars into several other subcategories, our CNN does not distinguish between strong and weak bars; anything flagged as a bar is thus treated as a bar. The unbarred training samples were randomly selected from the pool remaining NA10 samples. To avoid categorical biases, we used an equal number of barred and unbarred training samples. Altogether, we used 2,612 barred galaxy samples combined with an equal number unbarred samples for a total of 5,224 unique samples. We increased the number of samples through data augmentation. Data augmentation is a general technique for artificially boosting the size of training data through transforming and manipulating the input images, ultimately improving a CNNs ability to generalise (see \citealt{shorten2019} for a review). In following with the methodology of C21, the initial resolution of the training images is $110 \times 110$ pixels. Each image is cropped five times (one from each corner, plus the centre) to obtain five $100 \times 100$ cropped images. These cropped images were then further augmented via flipping and $90^\circ$ rotations, for a combined factor of 40 increase in the number of images. Hence the final dataset consisted of 208,960 images, each of size $100 \times 100$ pixels, and each linearly normalised such that every pixel has a floating point value between 0 and 1.

We used a similar procedure as in C21 for developing the new bar CNN. The overall training data was partitioned into two independent sets - one for training the CNN, the other for testing it - according to an 80\%:20\% split. We used the Python machine learning library \textsc{Keras} \citep{chollet2015}, running on \textsc{TensorFlow} \citep{abadi2016}, a general-purpose framework for machine learning. The networks were trained on a Nvidia RTX 1080Ti on ICRAR's \text{Pleiades} cluster. Figure \ref{fig:confmat} shows the confusion matrix for the bar CNN. The overall accuracy of the network is 83\%, which compares well with the accuracies of the models developed in C21. Unbarred galaxies are classified with a slightly higher accuracy of 84\%, compared to the 82\% accuracy of barred galaxies.

\section{Results}

All 35,082 images were independently classified by our CNNs as either barred or unbarred, and as either E, S0, Sp or IrrM. Figure \ref{fig:exbars} displays a random selection of barred galaxies that the bar CNN classified with very high confidence, i.e. $P(\text{bar}) > 0.95$. All the images in Figure \ref{fig:exbars} are the exact 100x100 images that comprised the input into the CNN. This is true for all future example illustrations. Among the provided examples are barred spiral galaxies, as well as strong bars within otherwise smooth discs.

\subsection{Bar Fraction}

We define the overall bar fraction as a function of redshift as simply the fraction of barred galaxies for all galaxies at a given redshift snapshot. Observations have concluded that the bar fraction generally decreases with increasing redshift \citep{sheth2008,masters2011,melvin2014}. It is known that the bar fraction varies strongly with stellar mass \citep{cameron2010,erwin2018}. In particular, \citet{nair2010a} found that a decrease in the bar fraction for higher mass samples, yet other studies (e.g. \citealt{melvin2014}) found the bar fraction increases with stellar mass. To analyse the impact of stellar mass on the bar fraction, we define three different mass ranges as follows: low-mass, from $10 \leq \log(M_\star/M_\odot) < 10.5$, intermediate-mass for $10.5 \leq \log(M_\star/M_\odot) < 11$, and lastly high-mass for $\log(M_\star/M_\odot) > 11$.

\begin{figure}
\centering
\includegraphics[scale=0.34]{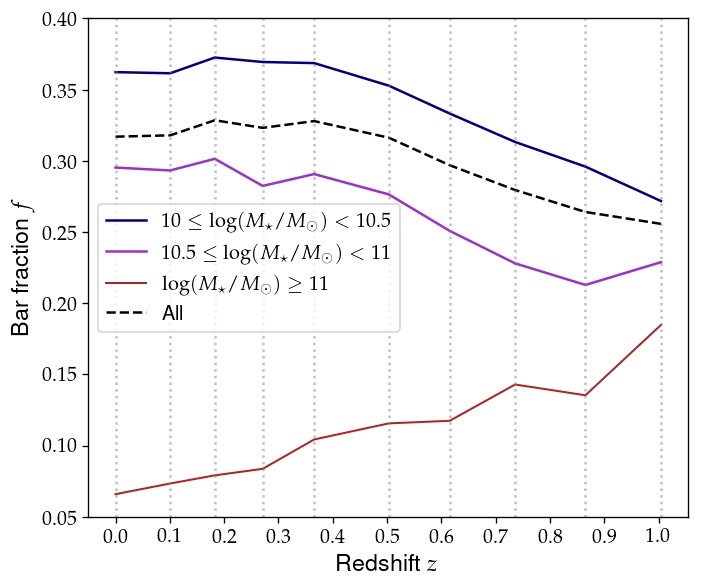}
\caption{The overall bar fractions as functions of redshift for low, intermediate and high mass ranges. The dotted line denotes the bar fraction for all samples regardless of mass.}
\label{fig:barf-bymass}
\end{figure}

\begin{figure}
\centering
\includegraphics[scale=0.52]{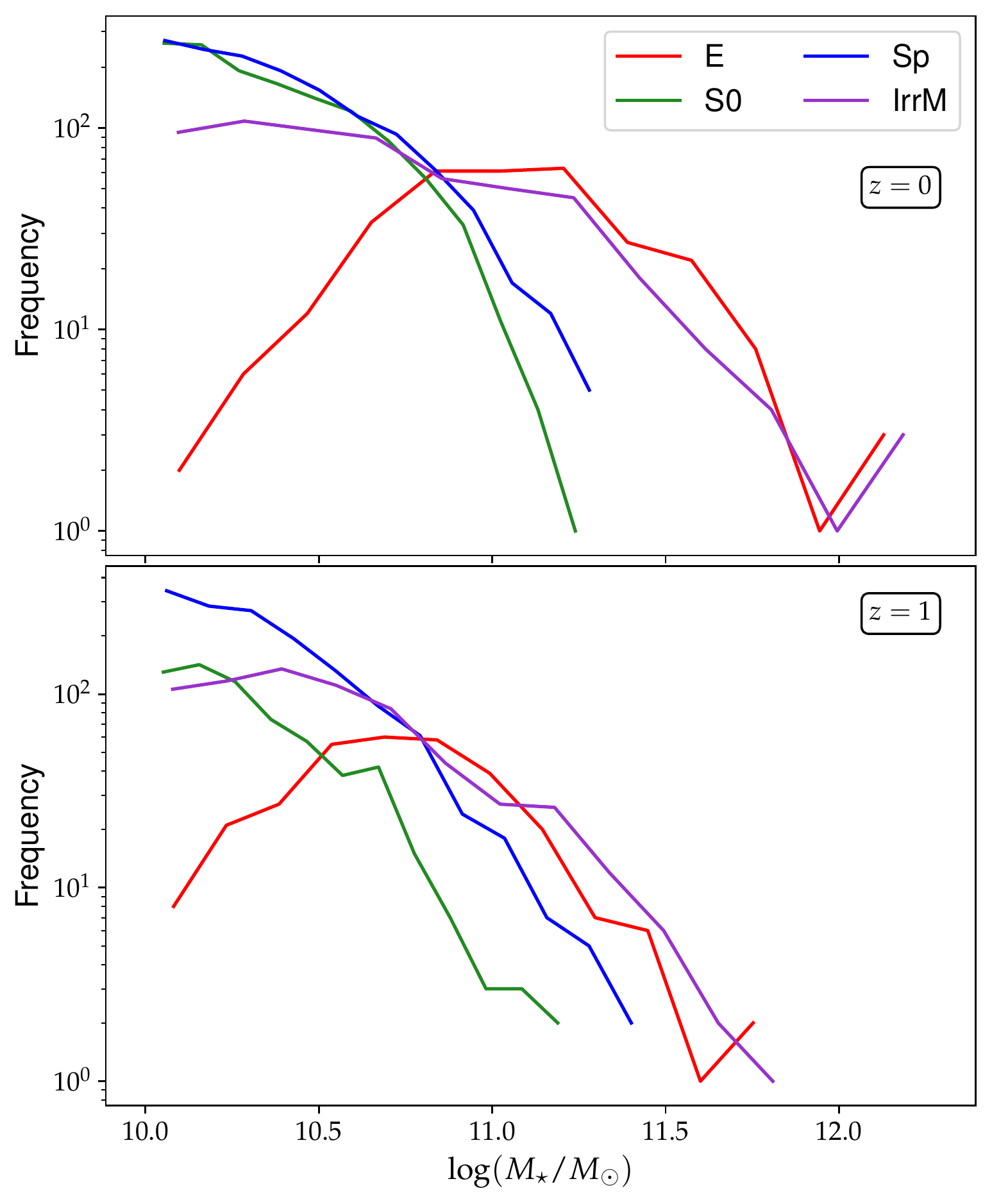}
\caption{Frequency of each of the four morphological types in terms of stellar mass $\log(M_\star/M_\odot)$ for $z=0$ (top panel) and $z=1$ (bottom panel).}
\label{fig:mtypefreq}
\end{figure}

Figure \ref{fig:barf-bymass} shows the overall bar fraction,
i.e. the bar fraction across all morphological types, for all three mass regimes, in addition to the overall bar fraction for all samples. This overall bar fraction remains at a roughly constant 32-34\% between $z=0$ and $z=0.5$, with a peak at about $z=0.366$. Beyond $z=0.5$, the bar fraction decreases to approximately 26\% at $z=1$. As for the separate mass ranges, we find that the bar fraction is highest in low-mass samples ($10^{10}$-$10^{10.5} M_\odot$), starting at around 37\% for $z = 0$ and decreasing to approximately 27\% for $z=1$. The intermediate-mass bar fraction decreases from just under 30\% at $z=0$ to 23\% at $z=1$, though notably it reaches a minimum of around 21\% at $z = 0.87$. The bar fraction in high-mass samples is considerably smaller, but actually increases with higher redshift to a final value of 18\% at $z=1$. In other words, bars in the most massive galaxies are less common now than they were in previous epochs. This suggests that the growth of new bars from $z=1$ to the present epoch is largely confined to low and intermediate-mass samples. Importantly, the number of galaxies above $10^{11} M_\odot$ is substantially smaller than that in the other two mass ranges; at $z=0$, just over 20 galaxies with $M_\odot \geq 10^{11}$ are barred, whilst over 800 in the low mass range are barred. As such, the high mass range is subject to low number statistics.
It is also worth noting the difference in morphologies across each mass range. Figure \ref{fig:mtypefreq} shows that, in general, the high mass range predominantly consists of elliptical galaxies, while disc galaxies are more common at intermediate and low masses. We note that the bar fraction in all three mass ranges appears to converge at higher redshifts. This is since the mass distributions for each morphological type are more similar at $z=1$. As the simulation progresses from $z=1$ to $z=0$, the high mass sample becomes more dominated by ellipticals, hence the reduced bar fraction at $z=0$ for this mass range as seen in Figure \ref{fig:barf-bymass}. Since the low and intermediate mass range consists mostly of discs, this hints that the growth in the bar fraction in these mass ranges is confined to disc morphologies. We will discuss morphology in more detail in the next section.

\begin{figure}
\centering
\includegraphics[scale=0.51]{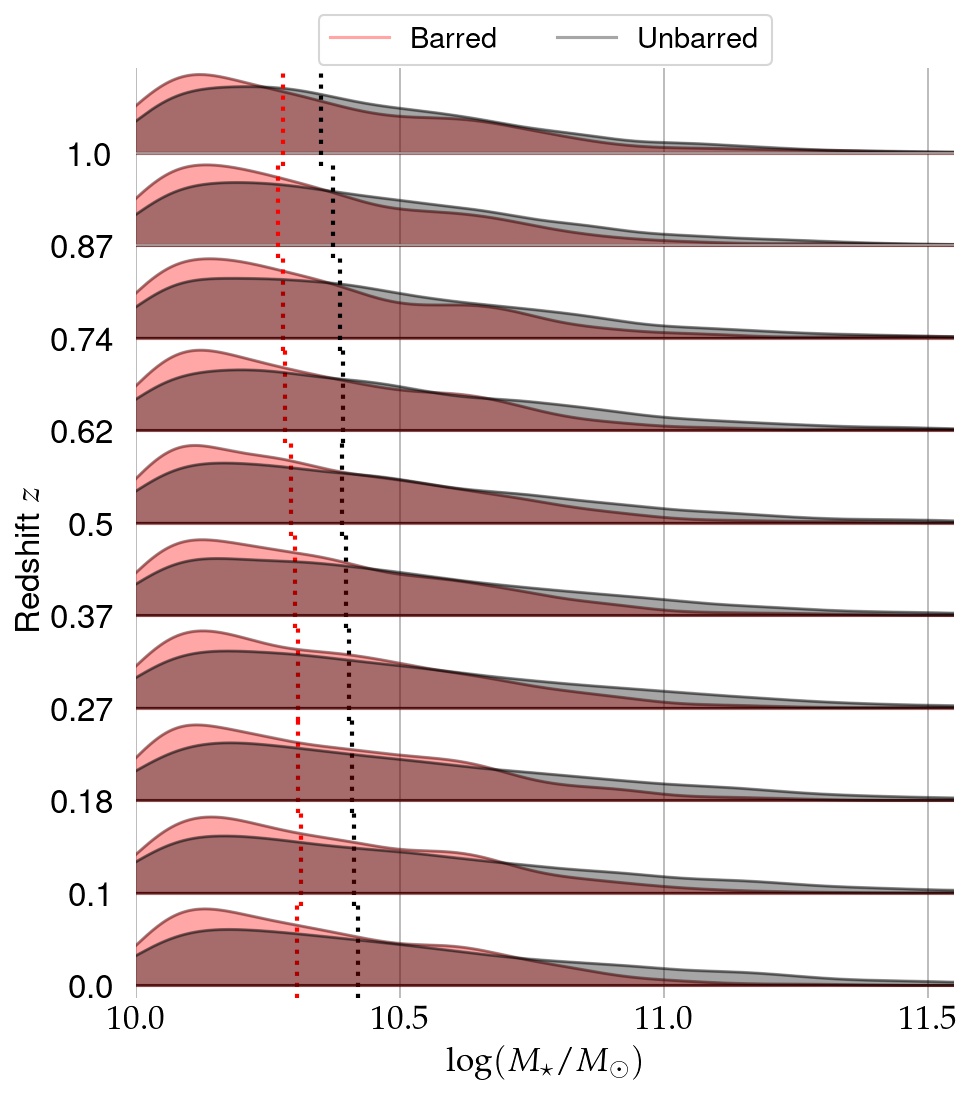}
\caption{Ridgeplot of the distributions of stellar masses for barred (red shading) and unbarred (black shading) galaxies, in units of $\log (M_\star / M_\odot)$, from $z=1$ to $z=0$. The dotted red and black lines denote the median stellar mass for barred and unbarred galaxies respectively.}
\label{fig:ridgeplotmass}
\end{figure}

Given the overall increase in bar fraction over time, it is worthwhile examining the change in overall mass distribution for barred and unbarred samples as a function of redshift, i.e. Figure \ref{fig:ridgeplotmass}. The median stellar mass for barred galaxies is roughly constant, with only a slight increase over time from $10^{10.28} M_\odot$ at $z=1$ to $10^{10.30} M_\odot$ at $z=0$. Unbarred samples instead show a slightly more substantial increase in stellar mass over time, with the median stellar mass of an unbarred galaxy increasing from $10^{10.34} M_\odot$ to $10^{10.42} M_\odot$. The larger mass growth for unbarred galaxies is consistent with the hierarchical growth of early type galaxies. This is also consistent with the increase in the mass distribution of ellipticals from $z=1$ to $z=0$, as seen in Figure \ref{fig:mtypefreq}. Furthermore, Figure \ref{fig:barf-bymass} shows that the bar fraction for high mass samples decreases slightly. This is consistent with the increase in the tail of the mass distribution for unbarred samples as $z$ decreases, and that the tail of the mass distribution for barred samples remains constant for all $z$.

When it comes to comparing these results with observed bar fractions, it is important to be mindful of the different sample selection criteria used, as well as the method used to detect the bar (see \citealt{lee2019}). As for mass ranges, \citet{melvin2014} used a similar low/intermediate/high mass range approach but instead found that low mass galaxies had considerably lower bar fractions than high mass galaxies. \citet{erwin2018} obtained the opposite, showing a sharp decline in bar fraction for masses higher than $\log(M_\star/M_\odot) \approx 10$. In terms of the values at the present redshift $z=0$, the overall fraction of 32\% compares well with established results based on observations, e.g. 30\% \citep{nair2010a}, 29.4\% \citep{masters2011}, 30.4\% \citep{lee2012}.

\subsection{Morphology}

\begin{figure}
\centering
\includegraphics[scale=0.54]{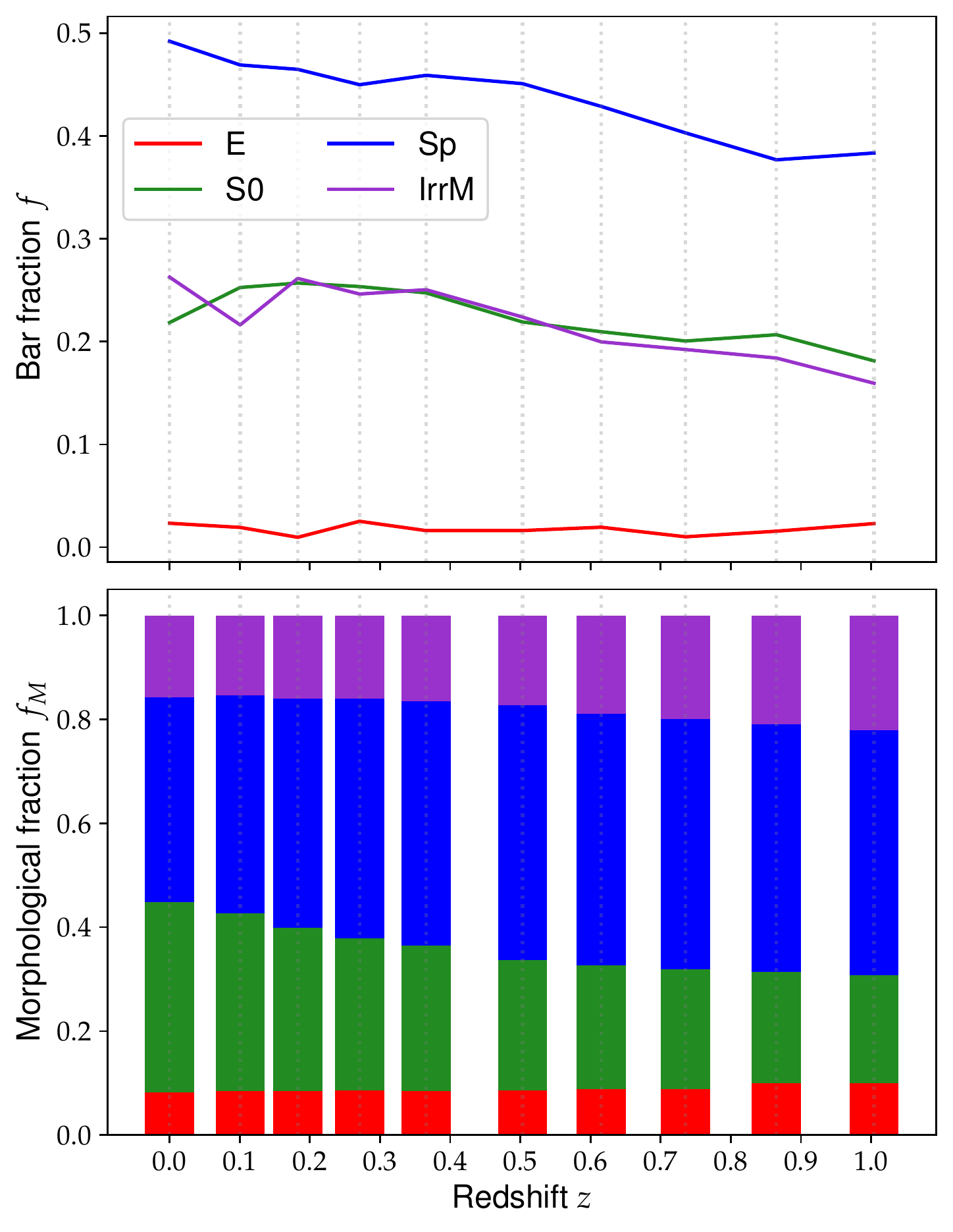}
\caption{(Top) The overall bar fraction, for samples classified as either of the four morphological types, as a function of redshift. (Bottom) Stacked bar plot of the fraction of samples in each morphological type at each snapshot.}
\label{fig:barf-typemorf}
\end{figure}
\begin{figure*}
\centering
\includegraphics[scale=0.48]{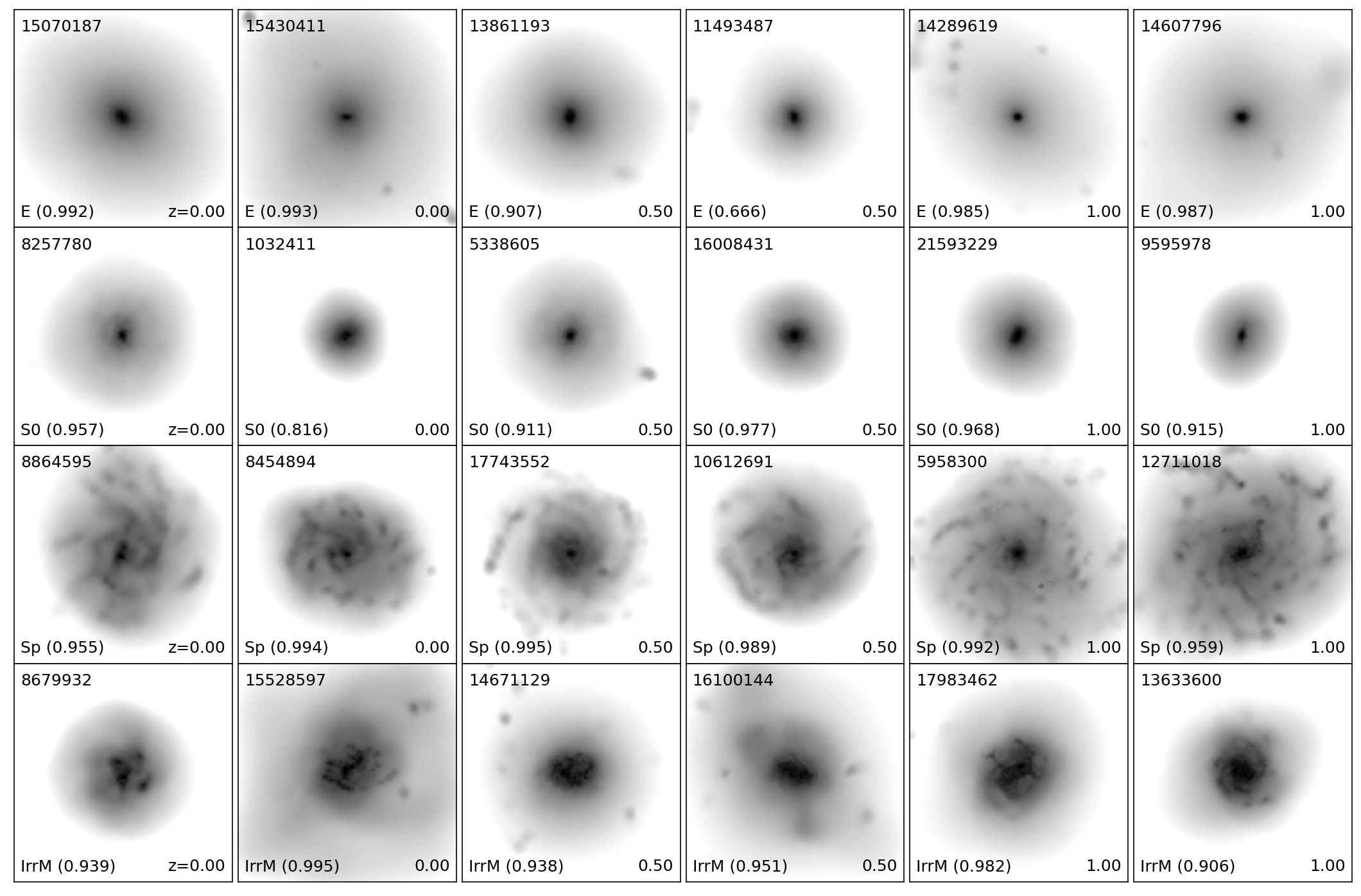}
\caption{Example of randomly selected unbarred samples with high classification confidences for each of the four main morphological types.}
\label{fig:ex4way}
\end{figure*}

Bars are most commonly observed in spiral galaxies. Estimates for the bar fraction in spirals range from 55\% \citep{aguerri2009} to as high as 70\% \citep{eskridge2000,sheth2008}. To investigate the bar fraction in different morphological types, we further classified all 35,082 with the 4-way CNN from C21. Figure \ref{fig:barf-typemorf} shows the bar fraction by morphological type for all galaxies regardless of stellar mass, in addition to the fraction of each morphology, as functions of redshift. We see that all morphologies, except for ellipticals, exhibit a decline in bar fraction over increasing redshift. As seen in observations (e.g. \citealt{aguerri2009,barway2011}), the bar fraction in spirals is higher than in all the other morphological classes, with a value of 49\% at $z=0$. Both lenticulars and irregular / miscellaneous samples have similar bar fractions, while the bar fraction in ellipticals remains very low (<5\%) throughout all snapshots, with a value of 3\% at $z=0$.

Regarding how common each morphological type is, we see in the bottom panel of Figure \ref{fig:barf-typemorf} that the fraction of spiral galaxies rises with redshift from around 40\% at $z=0$, peaks at 49\% at $z = 0.5$, then decreases slightly to 47\% at $z=1$. The rise in the fraction of S0s is considerably stronger at lower redshifts, with growth accelerating for $z < 0.5$.
The fraction of ellipticals remains relatively flat throughout all snapshots, having dropped slightly from a peak of approximately 10\% at $z=1$ to 8\% at $z=0$. Although this slight reduction is not consistent with observations \citep{bremer2018}, the ellipticals nevertheless become more massive over time. Irregulars similarly peak at $z=1$, becoming less common over time to just over 15\% at $z=0$.
The reduction in spirals from $z=0.5$ to $z=0$, and corresponding increase in S0s, is not reflected by any significant change in the evolution of the bar fraction;
both the S0 and Sp bar fraction continue to increase slightly within redshift range. Since the overall fraction of irregulars and ellipticals remains constant for $z < 0.5$, galaxies are mostly transitioning from spirals to S0s.
As the bar fraction in both S0s and spirals increases with decreasing $z$, this suggests that both barred spirals and unbarred spirals are transitioning to S0.
Mechanisms known to facilitate such transitions include gas stripping, tidal interactions and mergers, and the fading of the disc and/or spiral arms \citep{barway2009,bekki2011,borlaff2014,querejeta2015,fraser-mckelvie2018,rizzo2018}. We will discuss possible avenues for both bar creation and destruction in more detail in Section \S 4.

We further investigated the bar fractions across the different mass ranges in each of the four morphologies. We found, at $z=0$, that the bar fraction in S0s is highest in the intermediate mass range (33\%), while the low mass range S0 bar fraction (19\%) is lower than the overall S0 bar fraction (22\%). This is only observed for S0s: for all other morphologies, the bar fraction is highest in the low mass range. Our intermediate-mass range spirals have a bar fraction of around 41\% at $z=0$, comparing well with the value of 40\% that \citet{algorry2017} obtained for a selection of 296 \textsc{EAGLE} discs in the mass range $10.6 < \log(M_\star/M_\odot) < 11$. We note that \citet{algorry2017} used a Fourier analysis method to identify bars under a much stricter selection criteria, examining only disc galaxies. Our CNN method is applied to all galaxies within each snapshot, subject to the mass limits as outlined in \S 2.2, and so this approach allows us to examine both bar fractions by morphological type, as well as the overall bar fraction across all types. We further note that there is significant variation in the bar fractions obtained from simulations, e.g. 9\% to 7\% have been obtained from Illustris-1 and NewHorizon \citep{zhou2020,reddish2021}, to 55\% in TNG100 \citep{zhao2020}.
\begin{figure*}
\centering
\includegraphics[scale=0.53]{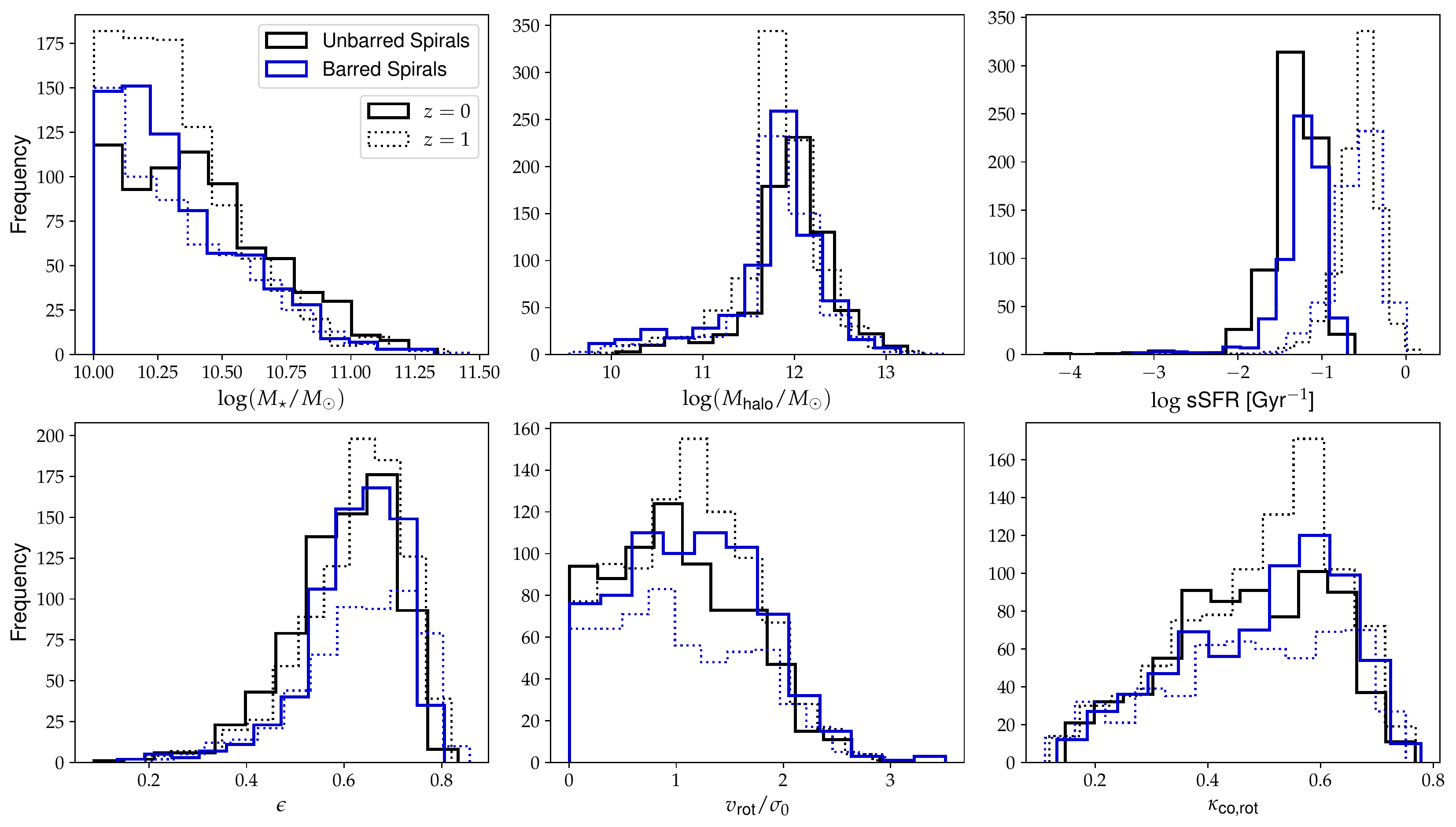}
\caption{Frequency histograms for barred (blue) and unbarred (black) spiral galaxies for the stellar mass $M_\star$, dark matter mass $M_{\text{halo}}$, specific star formation rate, ellipticity $\epsilon$, rotation to dispersion ratio $v_{\text{rot}}/\sigma_0$ and the corotation parameter $\kappa_{\text{co,rot}}$. The solid and dotted lines indicate the distributions at $z=0$ and $z=1$ respectively. Kinematic properties are calculated from all stellar particles with spherical radius $r < 30$ proper kpc.}
\label{fig:hist-properties-sp}
\end{figure*}
\begin{figure*}
\centering
\includegraphics[scale=0.53]{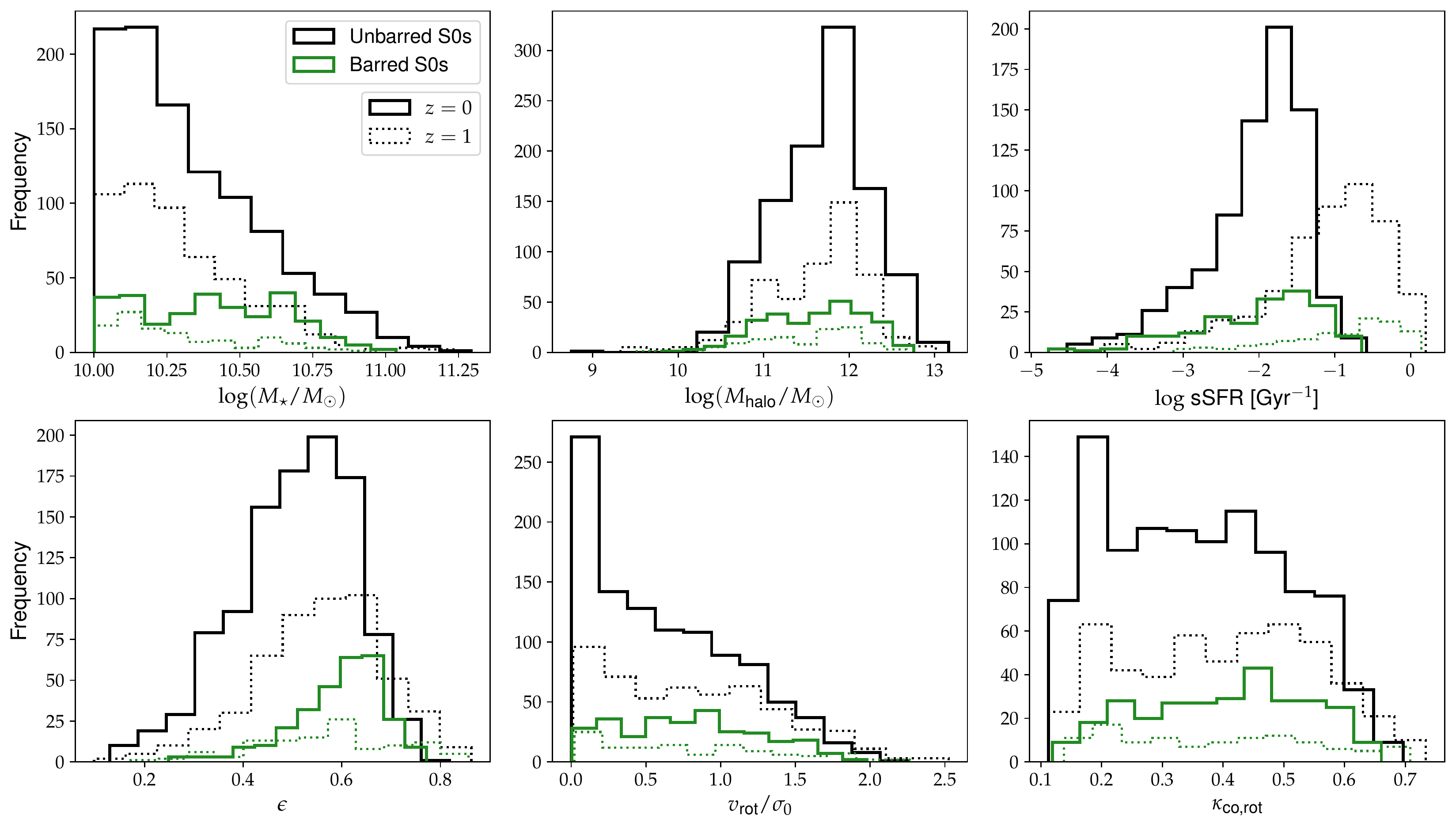}
\caption{Frequency histograms for barred (green) and unbarred (black) lenticular galaxies, for the properties listed as in Figure \ref{fig:hist-properties-sp}. The solid and dotted lines indicate the distributions at $z=0$ and $z=1$ respectively.}
\label{fig:hist-properties-s0}
\end{figure*}

Figure \ref{fig:ex4way} shows a random selection of high-confidence galaxies classified in each of the four morphological classes for $z=0$, $z=0.5$ and $z=1$. Visually inspecting a random sample of galaxies is a quick way to verify that the morphological classifications obtained with the CNN are appropriate, or if there are deeper issues with the models. As expected, ellipticals are more spheroidal in nature while lenticulars have a more prominent disk component, although in general it can be difficult to distinguish between ellipticals and lenticulars based solely on an image \citep{blanton2009}. Spirals are clearly distinguished, with the irregular / miscellaneous class examples clearly not fitting into any of the other three categories. We note that the miscellaneous samples from the NA10 dataset that the 4-way CNN was trained on contained several interacting galaxies, in addition to those with odd, ill-defined or clumpy shapes. This is likely why many galaxies with strong, turbulent gas flows are classified as irregular, even though they have a relatively symmetric and regular structure.

\section{Discussion}

\subsection{Properties of Barred Galaxies}

It is well known that galaxy bars have profound effects on the evolution of galaxies \citep{athanassoula2005,vera2016,fraser-mckelvie2020,geron2021}. To determine whether there is any clear impact, we examine the physical and kinematic properties of barred and unbarred galaxies, focusing on the two morphological types (S0 and Sp) in which bars were found to be most prevalent. For this, we utilise the \textsc{EAGLE} physical and kinematic properties described in \citet{mcalpine2016} and \citet{thob2019}. When it comes to disk galaxies, particularly barred galaxies, it is worth investigating the differences between spirals and lenticulars. Figures \ref{fig:hist-properties-sp} and \ref{fig:hist-properties-s0} display a series of histograms for various physical and kinematic properties for spirals and S0 galaxies respectively. These include stellar and dark matter masses, specific star formation rate, ellipticity (flattening) $\epsilon$, the ratio of rotational velocity to dispersion $v_{\text{rot}}/\sigma_0$, and the fraction of kinetic energy that is invested in corotation $\kappa_{\text{co,rot}}$ \citep{correa2017}. The kinematic properties are calculated based on all stellar particles with a spherical radius $r < 30$ proper kpc (or pkpc). Note that $\epsilon$ is defined as $1 - b/a$ where $a, b$ are the moduli of the major and minor axes of the ellipsoid corresponding to the spatial distribution of stars (see \citealt{thob2019} for details). Thus $\epsilon$ is inherently based on the 3D shape of the galaxy, and not just the ellipticity of the 2D face-on projection. We see in Figure \ref{fig:hist-properties-sp} that barred spirals tend to be less massive than unbarred spirals, and slightly more star forming. There is no clear distinction between barred and unbarred spirals in terms of ellipticity, however barred spirals have slightly higher values of $v_{\text{rot}}/\sigma_0$ and $\kappa_{\text{co,rot}}$. Lenticulars, shown in Figure \ref{fig:hist-properties-s0}, clearly have a different distribution in these parameters. In particular, there is a near-flat stellar mass distribution from around $10^{10}$ to $10^{10.75} M_\odot$ for barred S0s. There are also more intermediate-mass barred S0s at $z=0$ compared to $z=1$. Barred S0s also have a flatter spread in values of specific star formation compared to unbarred S0s. Unlike spirals, there is a clear distinction between barred and unbarred S0s in terms of ellipticity. Furthermore, the distributions of $\kappa_{\text{co,rot}}$ for S0s is considerably more uniform than that for spirals.
\begin{table}
\centering
\caption{Table of median values for physical and kinematic properties of barred and unbarred spirals and S0s for different stellar mass ranges.}
\label{tab:medprop}
\begin{tabular}{|c|c|c|c|c|c|c|}
\hline 
Type & Mass Range & Bar & $\log f_g$ & $\log$ sSFR & $\kappa_{\text{co,rot}}$ & $D/T$ \\ 
\ & $\log(M_\star/M_\odot)$ & \ & \ & Gyr$^{-1}$ & \ & \\
\hline 
S0 & 10-10.5 & Y & -1.052 & -1.578 & 0.439 & 0.554 \\ 
\hline 
S0 & 10-10.5 & N & -0.924 & -1.757 & 0.362 & 0.382 \\ 
\hline
S0 & >10.5 & Y & -0.848 & -2.682 & 0.386 & 0.535 \\ 
\hline 
S0 & >10.5 & N & -0.056 & -2.455 & 0.307 & 0.318 \\ 
\hline
Sp & 10-10.5 & Y & 0.157 & -1.157 & 0.524 & 0.593 \\ 
\hline 
Sp & 10-10.5 & N & 0.148 & -1.249 & 0.475 & 0.523 \\ 
\hline 
Sp & >10.5 & Y & 0.323 & -1.313 & 0.550 & 0.680 \\ 
\hline 
Sp & >10.5 & N & 0.385 & -1.380 & 0.516 & 0.606 \\ 
\hline 
\end{tabular} 
\end{table}

\begin{figure*}
\centering
\includegraphics[scale=0.5]{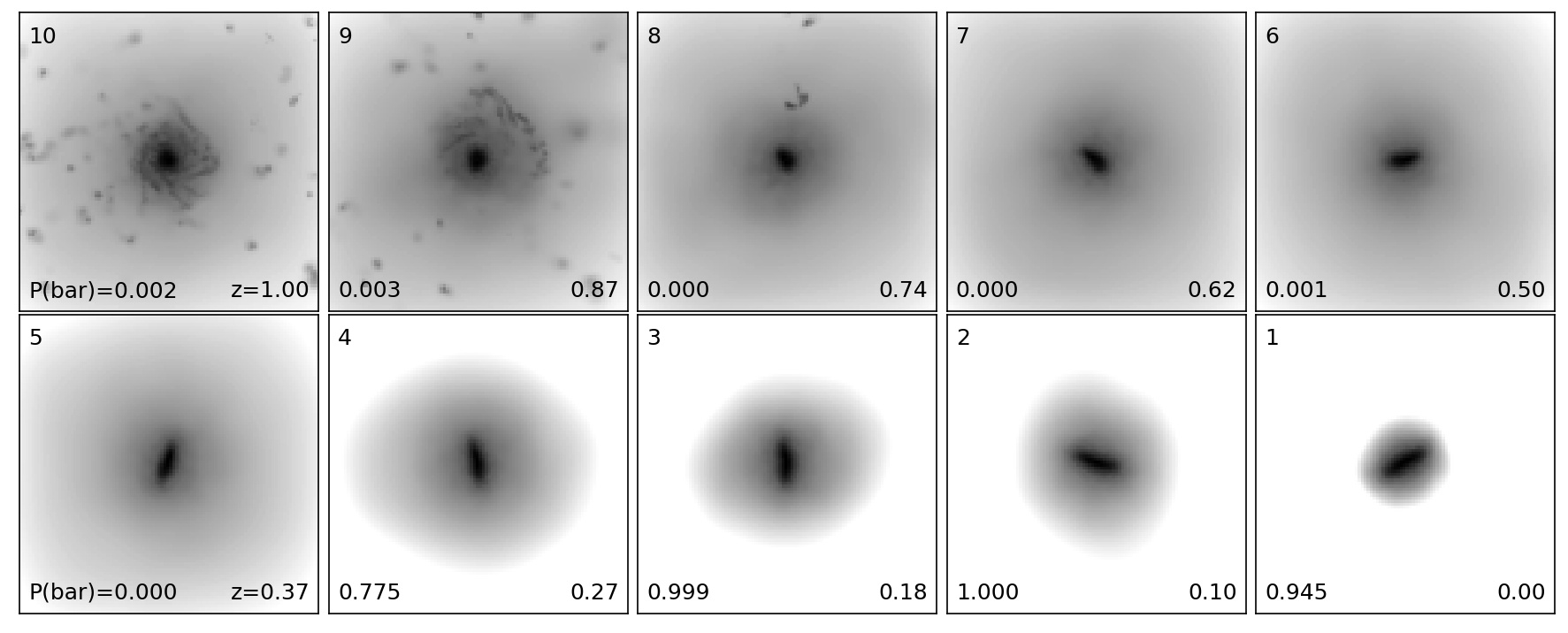}
\caption{Snapshots of a unique galaxy between $z=1$ to $z=0$ showing the evolution and growth of a galaxy bar.}
\label{fig:evolsample}
\end{figure*}

With this in mind, we investigate whether there are any clear properties that can distinguish barred spirals between barred S0s. Table \ref{tab:medprop} shows the median values for various physical and kinematic properties for barred and unbarred spirals at the present epoch $z=0$, and for a low and high mass range defined as less than and greater than $\log(M_\star/M_\odot) = 10.5$ respectively. As expected, gas fraction $f_g \equiv M_{\text{gas}} / M_\star$ is higher in spirals. However, while the gas fraction does not vary significantly among barred and unbarred spirals, for the case of S0s, barred S0s have lower gas fractions than their unbarred counterparts. This is considerably more apparent for the higher mass range. Star formation rates in S0s vary across mass range. High-mass barred S0s have lower star formation rates compared to unbarred S0s, while the reverse is true for low-mass S0s. Lower star formation rates in barred galaxies is consistent with bar-driven quenching and gas removal \citep{spinoso2017,fraser-mckelvie2020,geron2021}, but further analysis is required to investigate these mechanisms in detail. Across all types and mass ranges in Table \ref{tab:medprop}, barred galaxies have higher corotation parameters $\kappa_{\text{co,rot}}$, with values higher in spirals than for S0s, consistent with more energetic corotation.

Lastly, Table \ref{tab:medprop} also lists the values of $D/T$, i.e. the disc-to-total mass fraction. Barred spirals have higher median $D/T$ values compared to unbarred spirals, but the difference is significantly greater (up to 68\%) for S0 galaxies. That is, barred S0s are more strongly disc dominated compared to unbarred S0s. Conversely, unbarred S0s have relatively larger bulge components. It is known that bulges can hinder and even prevent bar formation in disc galaxies \citep{kataria2018}, and that the bar fraction is comparatively lower in galaxies with prominent bulges compared to galaxies with prominent discs \citep{barazza2008,aguerri2009}. It is also known that S0s have higher bulge fractions compared to spirals \citep{mishra2018}, that these are predominantly classical bulges compared to pseudobulges \citep{kormendy2004}, and that bulges are more common in higher mass S0s \citep{fisher2011}. This likely explains why the high mass S0s in Table \ref{tab:medprop} has the lowest median $D/T$ value of 0.318. We note in Table \ref{tab:medprop} that the $D/T$ values of barred S0s are similar to those of unbarred, low mass spirals.

\subsection{Bar Evolution}

Our results thus far have demonstrated that the bar fraction is in a state of flux, with bars comparatively more common at $z=0$ than they were at $z=1$. Given that \textsc{EAGLE} uses consistent IDs, and our CNN can consistently classify galaxies across snapshots, it is possible to track the evolution of every unique galaxy throughout all snapshots. In total, 2,146 unique galaxies are present for all ten snapshots between $z=1$ to $z=0$. These are the galaxies with $M_\star > 10^{10} M_\odot$ at $z = 1$ and do not merge with other galaxies. It is these galaxies that will form the basis of our analysis throughout the rest of this subsection.

Figure \ref{fig:evolsample} shows the evolution of the galaxy with ID ``10'' at $z=1$ through to $z=0$. This galaxy initially starts out as an unbarred spiral, and ends up evolving into a compact barred irregular. Of note here is that this galaxy is ungergoing tidal stripping from $z=0.27$ onwards, resulting in a more compact size at $z=0$. The bar grows slowly, and is first detected by the CNN with the image ID ``4'' (corresponding to $z=0.27$), after which the bar remains throughout the remaining snapshots. It is interesting to note that the CNN is yet to detect the bar at ID ``5'' ($z=0.37$) despite it being similar in length. It is likely that the more well defined disc in ID ``4'' -- resulting in a relatively more prominent bar -- led to the classification. 
Determining exactly when a bar is formed is thus difficult. The CNN classifications in Figure \ref{fig:evolsample} suggest the bar formed between $z=0.37$ and $z=0.27$, however it could be argued through visual inspection that the bar formed much earlier. It is also worth noting that the CNN do not distinguish between strong and weak bars, however it is likely that the CNN is biased towards strong bars as they are designed to extract the dominant features of a given image.

\begin{figure}
\centering
\includegraphics[scale=0.52]{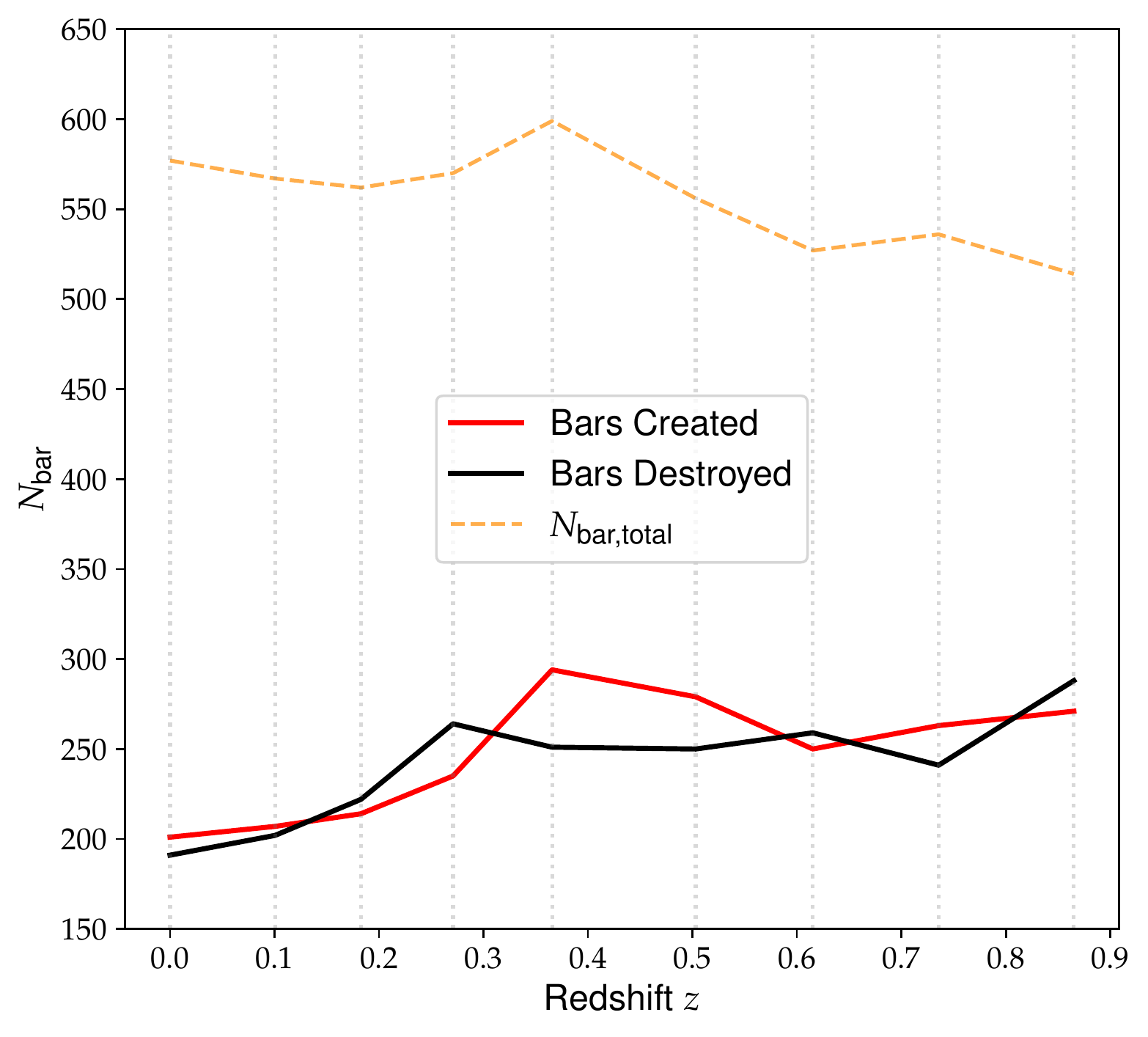}
\caption{Plot of the number of bars created and the number of bars destroyed between each snapshot (left vertical axis), along with the total number of bars $N_{\text{bar,total}}$ detected for each snapshot (right vertical axis), for all 2,146 galaxies present in all snapshots.}
\label{fig:barscreat}
\end{figure}

Nevertheless, we can still examine the creation and possible destruction of bars in \textsc{EAGLE}. We define bar creation and destruction as follows:
\begin{itemize}
\item Bar creation: The sample is classified as unbarred in the current snapshot $t$ and barred in the next snapshot $t+1$
\item Bar destruction: The sample is classified as barred in the current snapshot $t$ and unbarred in the next snapshot $t+1$
\end{itemize} Figure \ref{fig:barscreat} shows the number of bars created and destroyed between snapshots (note that, as per the above definitions, the initial snapshot at $z=1$ is excluded). We can see that there is a general downward trend in both the number of bars created and destroyed per snapshot. However, bar creation rises from $z=0.62$ to $z=0.36$, peaking at $z=0.36$. The difference between the number of bars created and the number of bars destroyed is also greatest within this redshift range, which can be thought of as an epoch of greatest net bar formation. Indeed, 50\% more bars are created at $z=0.36$ compared to at $z=0$, and from $z=0.62$ to $z=0.36$ the total number of bars increases by 20\%. At $z=0.27$ and $z=0.18$ more bars are being destroyed than created, leading to a drop in the total number of bars, before bars created once again outnumber bars destroyed in the final two snapshots. It is also curious to note that, solely for these 2,146 samples, the bar fraction is, on average, lower at around 26\% to 28\%. Given that the average stellar mass of these samples grows from $10^{10.56} M_\odot$ to $10^{10.81} M_\odot$ from $z=1$ to $z=0$, this is consistent with the lower overall fraction for intermediate mass samples as in Figure \ref{fig:barf-bymass}.

\begin{figure}
\centering
\includegraphics[scale=0.52]{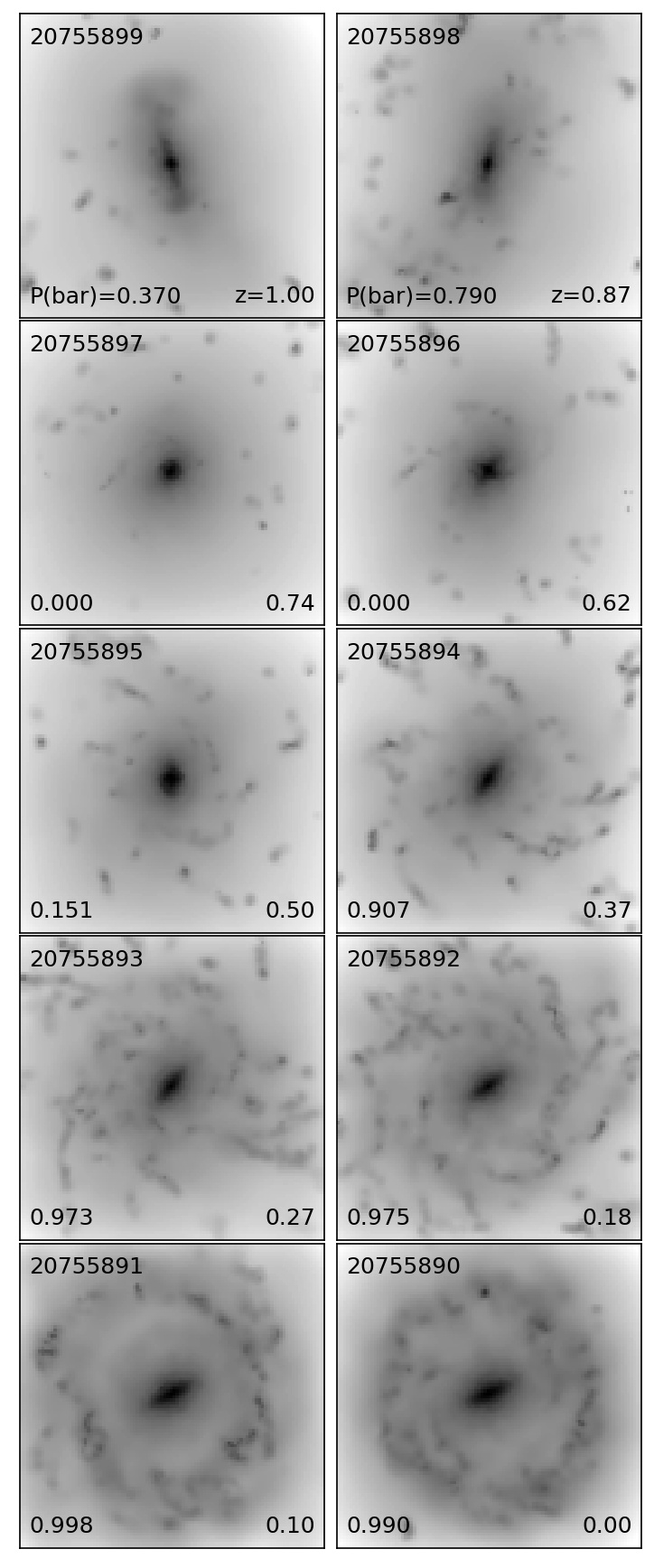}
\caption{Evolution of an example sample over all ten snapshots, illustrating both the destruction and regeneration of a bar.}
\label{fig:barregen}
\end{figure}

On average, 246 bars are created per snapshot and 240 are destroyed (roughly 11\% of all samples). It is inevitable that some of these transition events are caused by CNN misclassification. Given the maximum network accuracy of 82\% for barred galaxies in Figure \ref{fig:confmat}, this suggests that 18\% of detected bars may be erroneous, corresponding to roughly 100 bars per snapshot. Thus 20\% to 25\% of all bar creation and destruction events may be artificial, resulting instead from CNN misclassification. This is less than the 50\% change in bar creation events from $z=0.36$ to $z=0$, and the number of bars created and destroyed per snapshot is well above the level of misclassification. 
The trends in Figure \ref{fig:barscreat} nevertheless hint that the bar creation rate is not constant but also evolves with redshift. It has been posited that bars are transient features; simulations have shown that they are capable of being destroyed and regenerated, and that galaxies may well play host to more than one bar over their lifetime \citep{berentzen2004,bournaud2005}. We thus investigate the incidence of multiple bar creation and/or destruction events throughout the lifetime of each sample.

\begin{figure}
\centering
\includegraphics[scale=0.53]{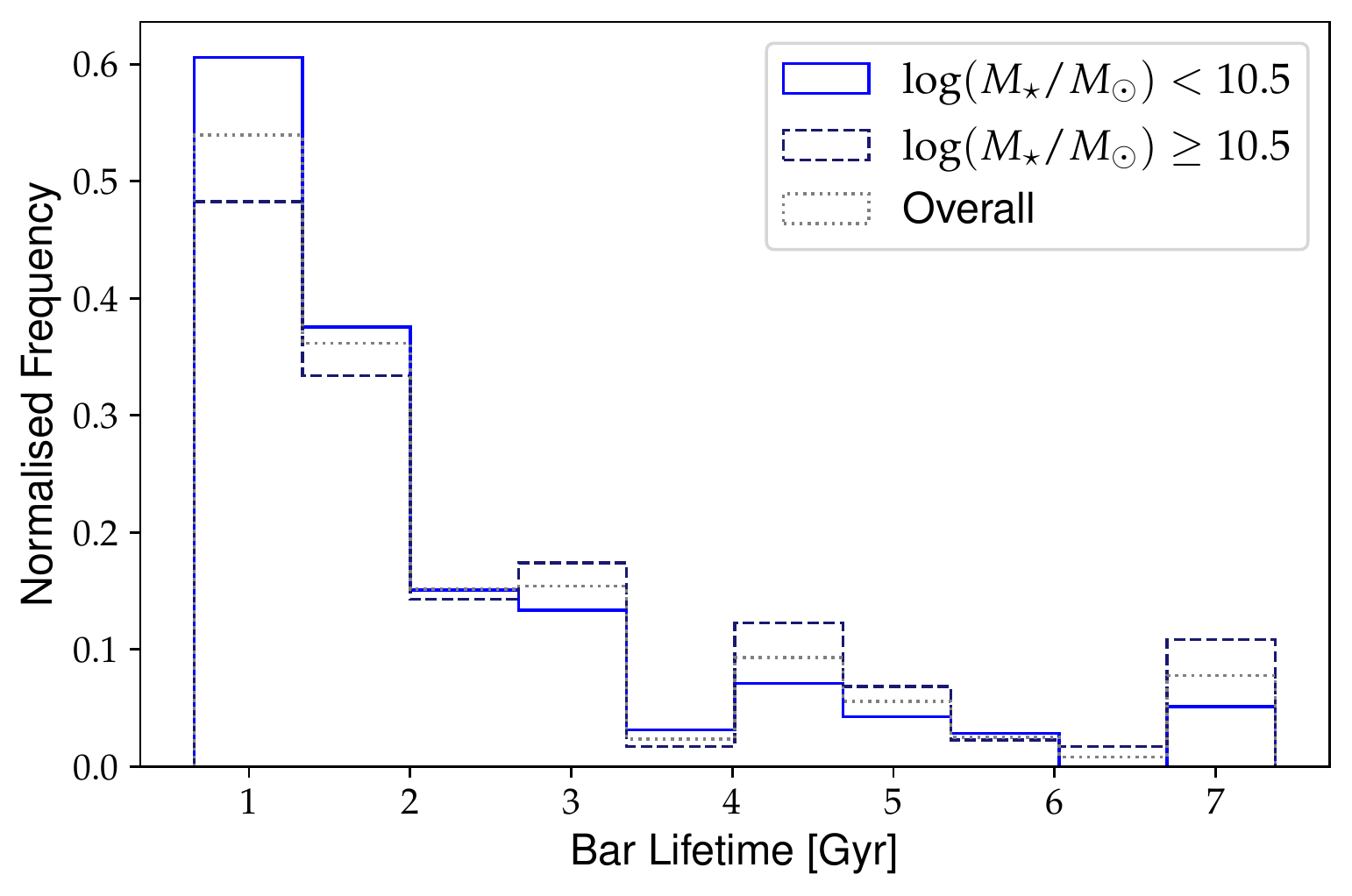}
\caption{Histogram of bar lifetimes for low-mass galaxies and high-mass galaxies.}
\label{fig:barlifetimes}
\end{figure}

Of the 2,146 samples, 35.1\% either did not form a bar or remained barred across all 10 snapshots, 35.8\% formed one bar, 22.1\% formed two bars and 7\% formed three or more bars, where bar formation is indicated via a transition from unbarred to barred in consecutive snapshots. Less than 1\% of samples remained barred throughout all ten snapshots. 65\% of all 2,146 samples experienced at least one bar destruction event, and 29\% experiencing two or more destruction events. Here bar destruction, like creation, is defined as a change in classification from barred to unbarred in consecutive snapshots. In total, 2,214 bars were created, 1,072 of which persisted for at least two consecutive snapshots.
This motivated the analysis of bar lifetimes, as determined by the difference in lookback times for the initial and final snapshots of the bar's existence (excluding snapshots at the very start and end of the simulation). Figure \ref{fig:barlifetimes} shows the distribution of bar lifetimes for low-mass and high-mass galaxies. The mean bar lifetime was found to be 2.24 Gyr, with a slightly shorter lifetime of 2.01 Gyr for low-mass galaxies, and 2.46 Gyr for high-mass galaxies. Note in this context, low mass galaxies are those with $\log(M_\star/M_\odot) < 10.5$ for all consecutive snapshots in which it is classified as barred, and high mass are those with $\log(M_\star/M_\odot) \geq 10.5$. The overall mean bar lifetime is slightly higher than the 1-2 Gyr lifetime found in simulations by \citet{bournaud2005}. It is extremely difficult to obtain estimates of bar lifetimes from observations, however analysis of accretions times such as in \citet{elmegreen2009}, have yielded similar ages.

It is important to note that this analysis is restricted only to galaxies present for all ten snapshots, which accounted for 54\% of all unique galaxies present. If this were not the case, the above fractions would be skewed towards one or less bar formation events due to the shorter timescales for galaxies that are lost to mergers. It is also worth noting that for the samples that undergo high numbers of creation and destruction events, these events are more likely to be artificial, having been caused by CNN misclassifications.
Similarly, the peak at around 1 Gyr in Figure \ref{fig:barlifetimes} is likely inflated due to CNN misclassifications, especially in cases of multiple, short-lived bars. Hence actual bar lifetimes may be higher than our value 2.24 Gyr.
Figure \ref{fig:barregen} gives an example of a galaxy that underwent two bar formation events: one between $z=1$ and $z=0.87$, and the other between $z=0.5$ and $z=0.37$). By examining the sequence of images, the initial bar is clearly destroyed, after which a new bar begins to gradually form, likely driven by gas accretion \citep{bournaud2002}.

\subsection{Mechanisms for Bar Formation and Destruction}

One major factor affecting the formation and evolution of galaxies is environment, which covers everything from ram pressure stripping, tidal interactions and fly-bys, to galaxy merging. Galaxy mergers play key roles in the growth and evolution of galaxies \citep{barnes1992,cole2002,conselice2014} and their dynamical evolution, especially through changes in angular momentum \citep{lagos2018} and bulge growth \citep{aguerri2001}. Mergers also affect star formation, gas flows and AGN activity \citep{perret2014,capelo2015,pearson2019}. Mergers may also play a role in the formation and evolution of galaxy bars. Previous simulations have shown that mergers can promote the formation of a bar \citep{peirani2009,cavanagh2020}, however they can also weaken existing bars and potentially destroy them \citep{ghosh2021}.  Another external mechanism responsible for inducing bar formation is that of tidal interactions, such as those induced by galaxy-galaxy interactions \citep{noguchi1987,elmegreen1990,miwa1998,moetazedian2017,peschken2019} and flybys \citep{lang2014,zana2018}. Bars can also form spontaneously via secular evolution as a result of global, non-axisymmetric instabilities  \citep{hohl1971,combes1981,sellwood1993,athanassoula2003,kim2016}. Altogether, these various mechanisms are also capable of destroying bars. This can occur by tidal stripping, gas infall and/or buckling due to vertical instabilities, \citep{raha1991,bournaud2005,martinez-valpuesta2006,lokas2019,xiang2021}, with some physical processes capable of reforming bars after their destruction \citep{bournaud2002,berentzen2004,bournaud2005}.

\begin{figure}
\centering
\includegraphics[scale=0.39]{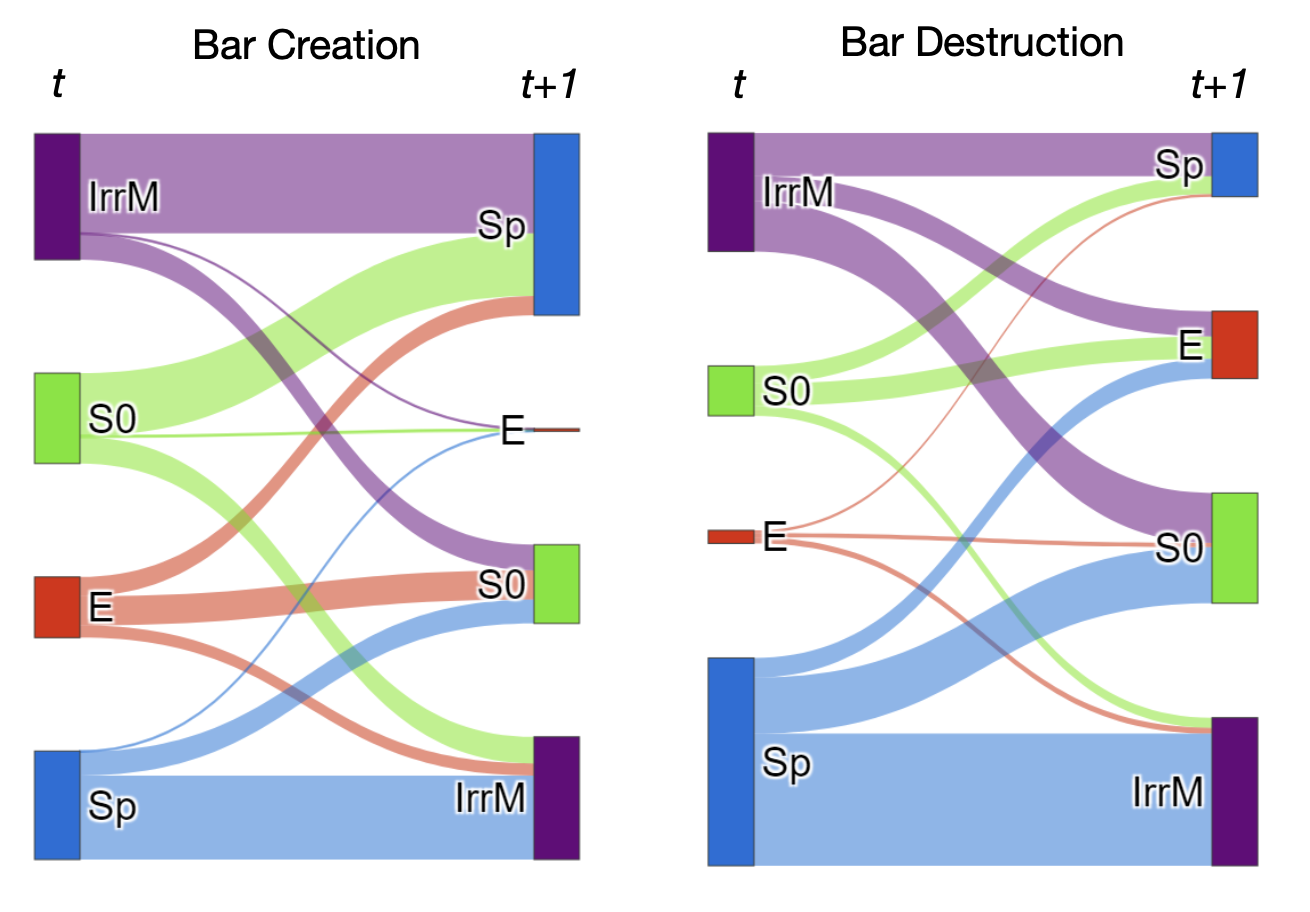}
\caption{Sankey diagrams summarising the changes in morphology between two snapshots as a result of bar creation (left) and bar destruction (right). This is restricted to only those samples for which the classified morphological type at snapshot $t+1$ differed to that at snapshot $t$, which comprises 24\% of the total bar creation events and 31\% of bar destruction events. The  size of the nodes and paths is proportional to the number of samples, and both plots are scaled to the same relative size.}
\label{fig:sankeybar}
\end{figure}

For all the 2,146 galaxies present for all snapshots, there were a grand total of 2,214 bar creation events, and 2,168 bar destruction events. Of the 2,214 bar creation events, 529 (24\%) also corresponded to a change in morphology (i.e. the morphological classification differed to that of the previous snapshot). Of the 2,168 bar destruction events, 673 (31\%) saw a change in morphology. Figure \ref{fig:sankeybar} illustrates the changes in morphology corresponding to bar creation and destruction events. Of these 529 morphological transitions corresponding to bar creation, the most common were unbarred irregulars/miscellaneous transitioning into barred spirals (29\%), unbarred spirals turning into barred irregulars (24\%) and unbarred S0s transitioning into barred spirals (18\%). Of the 673 morphological transitions corresponding to bar destruction, the most common change was from barred spirals to unbarred irregulars (37\%), followed by barred spirals to unbarred S0s (15\%) and barred irregulars to unbarred S0s (14\%). We note that further morphological transitions, especially those with very few samples, are more likely to correspond to CNN misclassifications.

The transition from unbarred irregulars to barred spirals is consistent with tidal interactions. That the reverse (barred spirals to unbarred irregulars) is the most common change in morphology associated with bar destruction is also consistent with the effects of tidal stripping and/or deformation, as well as major merging. The change from barred spiral to unbarred S0 is consistent with the faded spiral mechanism. Other transitions to unbarred S0s could be due to minor merging, or the destruction of the bar from bulge growth or gas infall\citep{vera2016,kruk2018}. Importantly, the majority of galaxies did not experience a change in morphology when forming a bar (76\%) or losing a bar (69\%).
\begin{figure}
\centering
\includegraphics[scale=0.54]{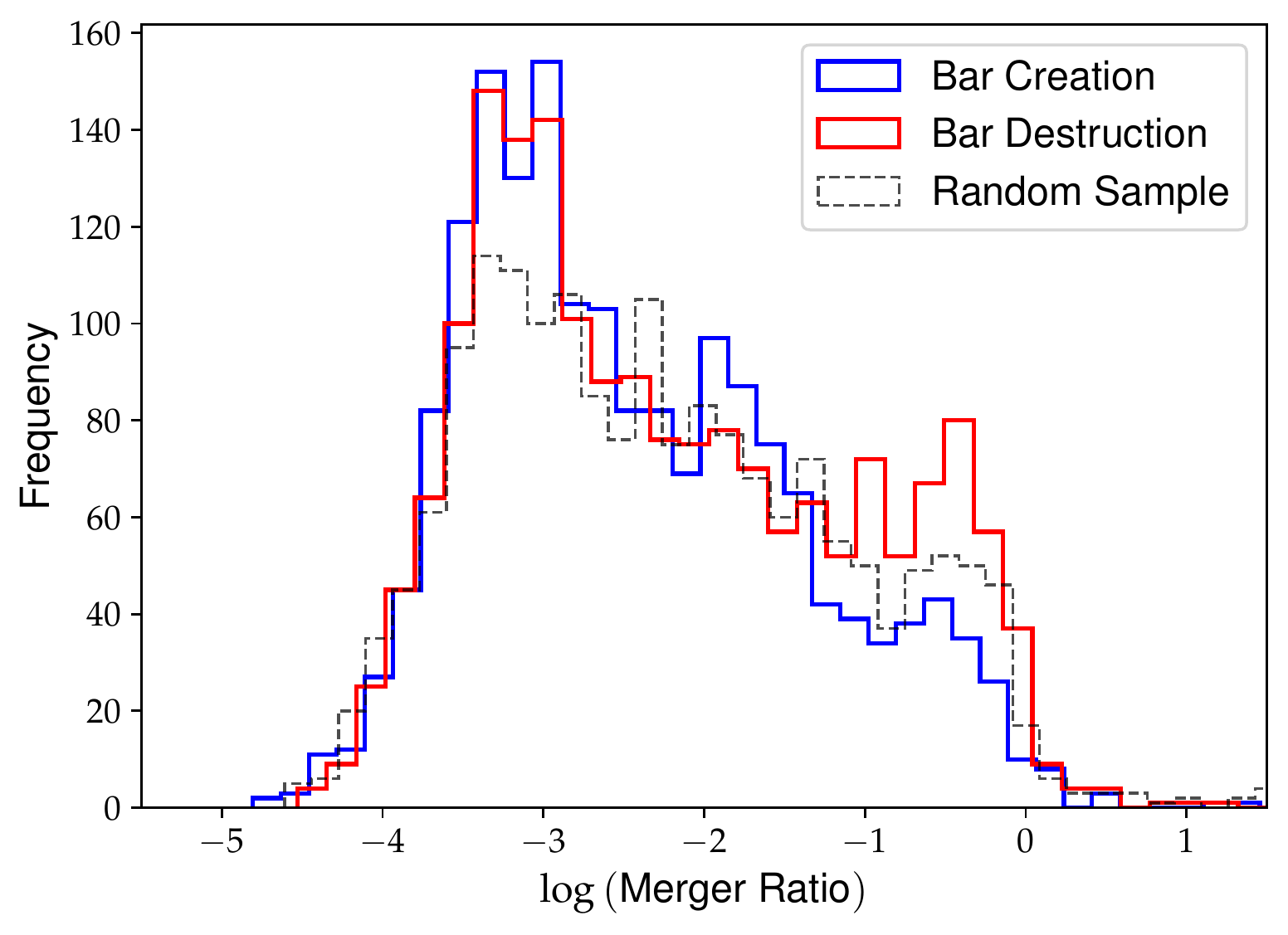}
\caption{Frequency distribution of merger ratios for mergers corresponding to times of bar creation (blue) and bar destruction (red). The dashed, black line corresponds to a random sample of merger ratios.}
\label{fig:mergers}
\end{figure}
\begin{figure*}
\centering
\includegraphics[scale=0.49]{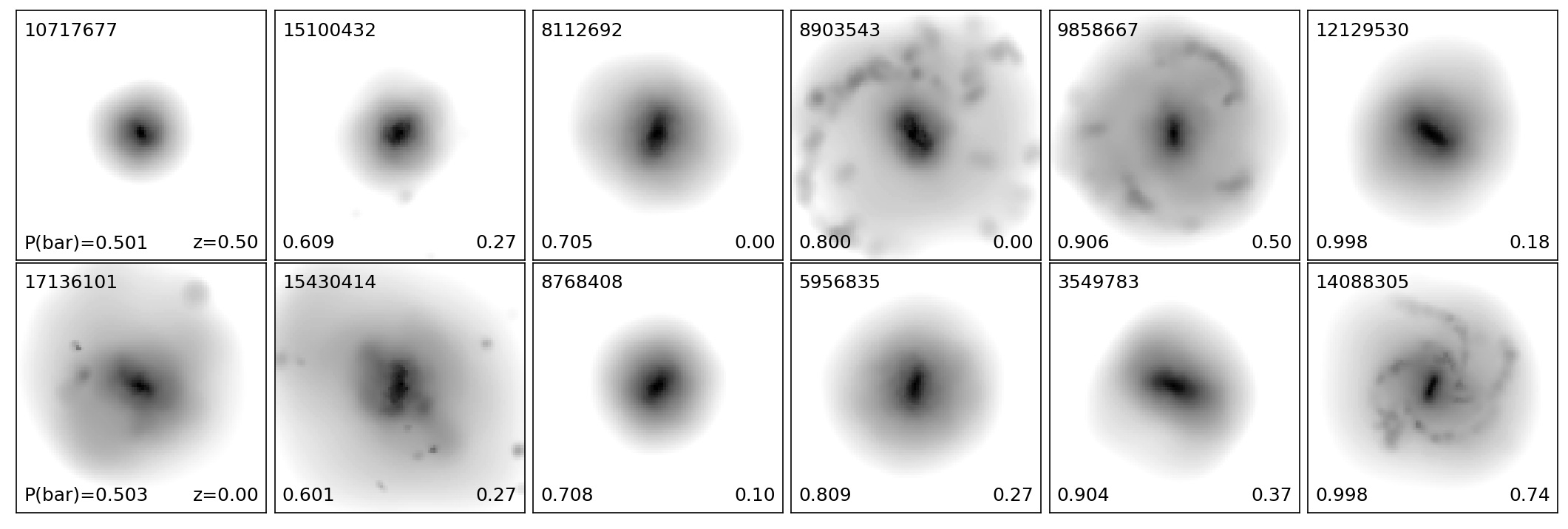}
\caption{A random selection of barred galaxies with classification confidences from 0.5 and 1.}
\label{fig:pbarproxy}
\end{figure*}

We can examine the effects of mergers on both bar creation and destruction through analysing the merger histories of each samples, and identifying the merger ratios of mergers linked to the snapshots that correspond to bar creation or destruction events. In Figure \ref{fig:mergers}, it can be seen that the merger ratios for bar creation and destruction share similar distributions, however there is a clear difference for the case of major merging ($\log$ merger ratio from -1 to 0). Here, major mergers are linked to up to twice the number of bar destruction events compared to bar creation events. Hence major merging is more strongly linked to bar destruction. This can also be seen by comparing the merger ratios for bar destruction to a random selection of merger ratios out of all mergers (black, dashed line in Figure \ref{fig:mergers}). This is consistent with previous N-body simulations from \citet{cavanagh2020}, which found that while mergers are capable of creating bars, major merging is far more likely to result in the destruction of a bar.

From Figure \ref{fig:mergers}, it can be seen that minor merging occurs with both creation and destruction, with a slight peak in bar creation at $\log(\text{ merger ratio }) = -2$, but all ratios are close to random. Smaller merger ratios (less than -3) are linked to accretion events. As for these accretion events, there is no noticeable difference in the distribution of merger ratios for bar destruction compared to bar creation. Crucially, however, is the fact that both distributions are higher than the random merger ratio sample. Although we cannot directly demonstrate an association, Figure \ref{fig:mergers} hints that accretion events are more strongly linked to both bar destruction and creation events, compared to a random sample of merger events. This comes as no surprise given the important role of accretion for the growth of bars \citep{bournaud2002,athanassoula2013}, as well as its capacity to weaken the bar and/or destroy it \citep{seo2019,ghosh2021}.

\subsection{Bar Strength}

As aforementioned in the previous subsection, our CNNs do not distinguish between strong and weak bars. The notion of bar strength is difficult to objectively define. Some studies have likened bar strength to the bar axis ratio \citep{martin1995}, to the ellipticity of the bar \citep{laurikainen2002} and others with the length of the bar \citep{hoyle2011}. Indeed, there are several approaches to quantitatively measure bar strength, which necessarily depend on how strength is defined. These range from measuring brightness or flux ratios of the bar in relation to the interbar region of the disc \citep{elmegreen1985,rozas1998}, to measuring the difference between the amplitudes of the $m = 0$ and $m = 2$ Fourier components \citep{aguerri1998,abraham2000,das2008,aguerri2009,garcia-gomez2017}. Some studies instead suggest weak and strong bars are extremes on continuum of bar strength \citep{geron2021}, with weaker bars sharing more in common with unbarred galaxies than barred galaxies \citep{abraham1999,cuomo2019}. In simulations, both \citet{algorry2017} and \citet{rosas-guevara2019} find that roughly half of the identified barred galaxies in \textsc{EAGLE} and \textsc{TNG100} respectively are strongly barred, with the other half weakly barred.

Bar strength can nevertheless be considered as an intuitive quality, however subjective that may be. Of the crowdsourced classifications of the Galaxy Zoo 2 project, strongly barred galaxies tend to correspond to those with majority vote-fractions ($p_{\text{bar}} > 0.5$), while weakly barred galaxies have lower vote fractions \citep{masters2012,skibba2012,willett2013}. We can attempt to see whether our CNN confidences, i.e. the probability that the image is barred, exhibit similar characteristics. Figure \ref{fig:pbarproxy} shows a random selection of galaxies with the probability of being barred, $P_{\text{bar}}$, varying from random (0.5) to near-certain. On visual inspection, the near-random classifications (0.5, 0.6) are indeed inconclusive; the examples in Figure \ref{fig:pbarproxy} are likely prolate bulges. Yet, as $P_{\text{bar}}$ increases, it can be argued that the presence of a bar becomes more clearer, in the sense that the bars itself become relatively more prominent. After all, a high classification confidence means that the CNN is more certain that it has detected the presence of a bar. It does not necessarily imply a stronger bar, only a more easily detectable bar.

We again stress that the CNNs do not directly distinguish between weak and strong bars, and therefore the probability of a sample being barred can only act as an effective proxy for bar strength (in an analogous manner to the Galaxy Zoo 2 vote fractions in \citealt{masters2012}). While it is impossible to delineate a hard threshold for defining a weak and strong bar in terms of our CNN classification confidences, we argue that these confidences are useful in a qualitative sense to distinguish bars that are well defined and prominent against those that are harder to detect the presence of (if any at all). This can be useful for identifying follow-up candidates in surveys for further visual classification, e.g. samples with low classification probabilities. Of course, these confidences are all ultimately calculated relative to the original training data used to train the CNN.

\subsection{CNN Evaluation}

With the continued development of large-scale cosmological simulations, in addition to future observational studies expected to probe billions of galaxies, there is a need for techniques for automated analysis in order to process such large amounts of data efficiently. The primary advantage of automated classification techniques, particularly CNNs, is their ability to generalise and scale, enabling otherwise intractable amounts of data to be processed within far more practical limits. The previous work by \citet{algorry2017} studied a selection of only 269 discs in \textsc{EAGLE}. Our CNN approach allows us to analyse all \textsc{EAGLE} galaxies within a single snapshot, across multiple snapshots, limited only by our mass range criteria. This approach is uniquely well suited to analyse bar creation and destruction, changes in morphology, as well as bar lifetimes.

CNNs are also highly versatile, and can be easily adapted for different classification tasks. While previous studies have applied CNNs trained on simulated data for use in classifying observational data \citep{ghosh2020,eriksen2020}, this work achieves the opposite. The reason we use synthetic SDSS-like g-band images is since these are the images that our CNNs are trained on. While it is possible to adapt a CNN to classify images in other morphological bands, this requires performing a model adaptation step in the form of transfer learning. Recent examples of this in astronomy include transfer learning between different surveys \citep{dominguezsanchez2019} and modifying general-purpose image classifiers \citep{ackermann2018}. For the purpose and scope of this work, restricting classifications to the g-band avoids the need for large-scale transfer learning.

\begin{figure}
\centering
\includegraphics[scale=0.54]{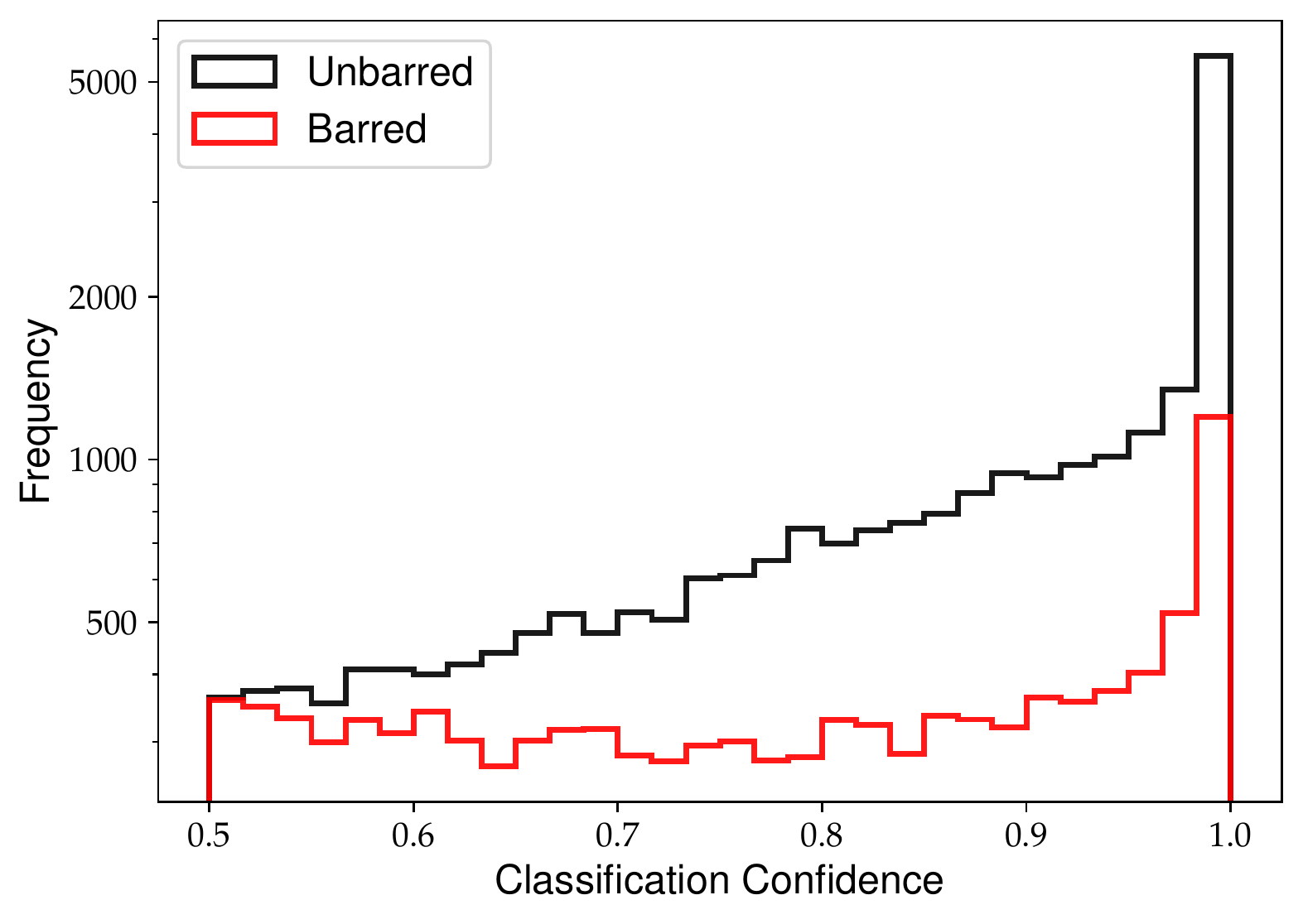}
\caption{Frequency histograms of the classification confidences – as classified by the bar CNN – for barred (red) and unbarred (black) samples. Note the scale is logarithmic.}
\label{fig:confidences}
\end{figure}

It is also important to note several limitations to the use of CNNs, especially for those reliant on labelled training data (i.e. supervised learning). The most crucial caveat is that supervised CNNs can only ever be as good as the data they are trained on. In the case of most large-scale, labelled observation datasets such as NA10 and GZ2, this upper limit is reflective on the limits of visual classification. CNNs are also prone to overfitting to the training the data, which can adversely affect performance when applied to broader data. Figure \ref{fig:confidences} shows the distribution of classification confidences for all barred and unbarred classifications. A long, smooth tail in both distributions is desirable compared to a narrow spike at high confidence, the latter of which is indicative of overfitting. The spread in classification confidences is also indicative of the difficulty in classifying bars.

\section{Conclusions}

In this work, we used CNNs trained to classify galaxy morphologies to study the evolution of barred galaxies in the \textsc{EAGLE} simulation. This was done by classifying a total of 35,082 synthetic g-band images of over 3,000 galaxy samples tracked over the redshift range $z=0$ to $z=1$. Our key conclusions are summarised follows:

\begin{enumerate}[(i)]
\item We found that the overall bar fraction in the g-band is roughly constant at 32\% to 34\% from $z=0$ to $z=0.5$, before exhibiting a general decline to 26\% at $z=1$. The values for the bar fractions at $z=0$ and beyond are generally consistent with previous observational studies. The bar fraction is highest in spiral galaxies (49\% at $z=0$), followed by irregular/miscellaneous samples (26\%), lenticulars (22\%) and ellipticals (3\%).
\item The bar fraction is higher in low-mass galaxies ($10^{10}M_\odot$ to $10^{10.5}M_\odot$. For S0 galaxies, bars are most common in intermediate-mass galaxies between $10^{10.5}$ and $10^{11} M_\odot$, while for all other morphological types, bars are most common in low-mass galaxies.
\item From $z=0.5$ to $z=0$, the fraction of spiral galaxies decreased from 49\% to under 40\%, while the fraction of S0 galaxies rose from 25\% to 37\%. The fraction of ellipticals and irregulars did not vary significantly. Although the bar fraction in spirals and S0s both rise with decreasing redshift, the bar fraction in S0s remains lower than that for spirals. This suggests that the growth in S0s may be driven by unbarred spirals transitioning into S0s, or that the transition from spiral to S0 tends to destroy the bar.
\item We have found that, in general, barred spirals are mostly similar to unbarred spirals. Barred spirals have greater values of $\kappa_{\text{co,rot}}$, slightly higher star formation rates, higher ellipticities and are less massive. Barred S0s have lower gas fractions than unbarred S0s, and also have lower star formation rates. Barred S0s also have significantly higher $D/T$ ratios, and are thus more strongly disc dominanted compared to their unbarred counterparts.
\item Bars have been shown to undergo episodes of creation, destruction and regeneration, with some galaxies forming multiple bars over the redshift range $z=1$ to $z=0$. Of the 2,146 samples present for all ten snapshots, 
35.8\% formed one bar, 22.1\% formed two bars and 7\% formed three or more bars.
\item We determined that, of the bars that persisted for at least two consecutive snapshots, the mean bar lifetime is 2.24 Gyr, with a slightly shorter mean lifetime of 2.01 Gyr for bars in low mass galaxies, and a longer mean lifetime of 2.46 Gyr for bars in high mass galaxies.
\item We have shown that rate of bar creation and bar destruction is not constant, with both declining for decreasing $z$. Bar creation peaks at $z=0.36$. The majority of bar creation and destruction events corresponded with no change in morphology. Of the events which did result in a change in morphology, the most common transition for bar creation was unbarred irregular/miscellaneous to barred spiral, and barred spiral to unbarred irregular/miscellaneous for bar destruction.
\item We have found that major merging is more strongly linked to bar destruction as opposed to bar creation. In addition, accretion is more strongly linked to both bar creation and destruction. This is since there are more instances of accretion linked to both a bar creation or destruction event compared to a random sample.
\end{enumerate}

\section*{Acknowledgements}

We wish to acknowledge and thank the anonymous referee for their constructive and insightful feedback that helped to improve the presentation of the paper. MKC acknowledges the support of the Australian Government Research Training Programme at the University of Western Australia. JP is supported by the Australian government through the Australian Research Council's Discovery Projects funding scheme (DP200102574). This work used the DiRAC Data Centric system at Durham University, operated by the Institute for Computational Cosmology on behalf of the STFC DiRAC HPC Facility {\url{www.dirac.ac.uk}}. This equipment was funded by BIS National E-infrastructure capital grant ST/K00042X/1, STFC capital grants ST/H008519/1 and ST/K00087X/1, STFC DiRAC Operations grant ST/K003267/1 and Durham University. DiRAC is part of the National E-Infrastructure.

\section*{Data Availability}

This project makes use of data from the \textsc{EAGLE} cosmological simulations. The data catalogues are publicly available at \url{https://virgodb.dur.ac.uk/}, and are described in \citet{mcalpine2016} and \citet{thob2019}. Specific data pertinent to this work, such as catalogues of classifications, will be made available upon reasonable request to the author.



\bibliographystyle{mnras}
\bibliography{mybib_final} 


\bsp	
\label{lastpage}
\end{document}